\documentclass[10pt,twoside,preprint2]{aastex}

\usepackage{natbib}
\usepackage{graphicx} 
\usepackage{subfigure}

\usepackage[scaled=1.00]{helvet}

\usepackage[T1]{fontenc}
\usepackage{textcomp}

\usepackage{ragged2e} 

\usepackage{sectsty}
\sectionfont{\normalsize}
\subsectionfont{\normalsize}

\usepackage{fancyhdr}

\usepackage{lscape}
\usepackage{indentfirst}
\usepackage{latexsym}
\usepackage{multirow}
\usepackage{tabls}
\usepackage{wrapfig}
\usepackage{slashbox}
\usepackage{longtable}

\newcommand{\mypar}{\subsubparagraph}

\makeatletter

\renewcommand\mypar{\@startsection{paragraph}{4}{0ex}{-3.25ex plus -1ex minus -0.2ex}{1.5ex plus 0.2ex}{\normalfont\normalsize\bfseries}}
\makeatother
\stepcounter{secnumdepth}
\stepcounter{tocdepth}

\renewcommand{\paragraph}{\subparagraph}

\begin{document}











\pagestyle{fancy}

\fancyhead{}
\fancyhf{}

\rightmark 
\leftmark

\fancyhead[LE,RO]{\bfseries\thepage}
\fancyhead[CE]{\small {\bf Bogdan Popescu} and {\bf M.M. Hanson 2010}}
\fancyhead[CO]{{\small MASSCLEAN{\fontfamily{ptm}\selectfont \textit{age}} -- Stellar Cluster Ages from Integrated Colors --  }}












\shorttitle{MASSCLEAN{\fontfamily{ptm}\selectfont \textit{age}} \\-- Stellar Cluster Ages --} 
\shortauthors{B. Popescu & M.M. Hanson}


\title{\LARGE{MASSCLEAN}{\fontfamily{ptm}\selectfont \textit{age}} \\-- {\Large } Stellar Cluster Ages from Integrated Colors --}

\normalsize

\author{Bogdan Popescu and M. M. Hanson}
\affil{Department of Physics, University of Cincinnati, PO Box 210011, Cincinnati, OH 45221-0011}
\email{popescb@mail.uc.edu, margaret.hanson@uc.edu}

\begin{abstract}

{\small  
We present the recently updated and expanded MASSCLEAN{\fontfamily{ptm}\selectfont \textit{colors}}, a database of 70 million Monte Carlo models selected to match the properties (metallicity, ages and masses) of stellar clusters found in the Large Magellanic Cloud (LMC). This database shows the rather extreme and non-Guassian distribution of integrated colors and magnitudes expected with different cluster age and mass and the enormous age degeneracy of integrated colors when mass is unknown.  This degeneracy could lead to catastrophic failures in estimating age with standard SSP models, particularly if most of the clusters are of intermediate or low mass, like in the LMC. Utilizing the MASSCLEAN{\fontfamily{ptm}\selectfont \textit{colors}} database, we have developed MASSCLEAN{\fontfamily{ptm}\selectfont \textit{age}}, a statistical inference package which assigns the most likely age and mass (solved simultaneously) to a cluster based only on its integrated broad-band photometric properties.  
Finally, we use MASSCLEAN{\fontfamily{ptm}\selectfont \textit{age}} to derive the age and mass of LMC clusters based on integrated photometry alone.  First we compare our cluster ages against those obtained for the same seven clusters using more accurate integrated spectroscopy. We find improved agreement with the integrated spectroscopy ages over the original photometric ages.
 A close examination of our results demonstrate the necessity of solving simultaneously for mass and age to reduce degeneracies in the cluster ages derived via integrated colors.  We then selected an additional subset of 30 photometric clusters with previously well constrained ages and independently derive their age using the MASSCLEAN{\fontfamily{ptm}\selectfont \textit{age}} with the same photometry with very good agreement. The  MASSCLEAN{\fontfamily{ptm}\selectfont \textit{age}} program is freely available under GNU General Public License.  
} 
\vskip 0.25cm


{\small Accepted for publication in {\it The Astrophysical Journal}}

\vskip 0.5cm

\end{abstract}

\keywords{galaxies: clusters: general --- methods: analytical --- open clusters and associations: general}

\section{Introduction}
Stellar clusters are among the most universally-useful entities studied by modern day astronomers.  Having a set of singly-aged, same-distance stars in an identifiable cluster provides stellar and galactic astronomers with powerful diagnostic tools to interpret our universe.   Of particular interest is the ability to use the ages and masses of constituent stellar clusters to track the star formation history and chemical evolution of a galaxy, the mass function of stellar clusters and their lifetime within differing galaxies (e.g.\ \citeauthor*{lamers2005} \citeyear{lamers2005}; \citeauthor*{piatti2009} \citeyear{piatti2009}; \citeauthor*{larsen2009} \citeyear{larsen2009}; \citeauthor*{mora2009} \citeyear{mora2009}; \citeauthor*{chandar2010a} \citeyear{chandar2010a}). However, such measures require deriving the ages and masses of stellar clusters when the constituent stars within those clusters are not individually resolved.  With distant stellar clusters, one has only the integrated properties to work from, typically in either broad band photometry or integrated spectroscopy, to estimate ages (e.g.\ \citeauthor*{searle} \citeyear{searle}; \citeauthor*{vdb} \citeyear{vdb}).  In such cases, we rely on sophisticated models of these star systems to derive the cluster properties (age, mass, chemistry) based on their observed, bulk properties. For this reason there exists a long history of effort interpreting integrated cluster properties through the use of these sophisticated models (e.g.\ \citeauthor*{girardi} \citeyear{girardi}; 
\citeauthor*{bruzual2003} \citeyear{bruzual2003}; \citeauthor*{cervino2009} \citeyear{cervino2009}; \citeauthor*{buzzoni} \citeyear{buzzoni}; \citeauthor*{chiosi} \citeyear{chiosi}; \citeauthor*{bruzual2001} \citeyear{bruzual2001}; \citeauthor*{bruzual2010} \citeyear{bruzual2010}; \citeauthor*{cervino2004} \citeyear{cervino2004}; \citeauthor*{cervino2006} \citeyear{cervino2006}; \citeauthor*{fagiolini2007} \citeyear{fagiolini2007}; \citeauthor*{lancon2000} \citeyear{lancon2000}; \citeauthor*{lancon2009} \citeyear{lancon2009}; \citeauthor*{fouesneau} \citeyear{fouesneau}; \citeauthor*{fouesneau2} \citeyear{fouesneau2}; \citeauthor*{brocato2000a} \citeyear{brocato2000a}; \citeauthor*{brocato2000b} \citeyear{brocato2000b}; 
\citeauthor*{cantiello} \citeyear{cantiello}; \citeauthor*{raimondo2005} \citeyear{raimondo2005}; \citeauthor*{raimondo2009} \citeyear{raimondo2009};
\citeauthor*{santos} \citeyear{santos}; \citeauthor*{gonzalez} \citeyear{gonzalez}; \citeauthor*{gonzalez2005} \citeyear{gonzalez2005}; \citeauthor*{gonzalez2010} \citeyear{gonzalez2010}; \citeauthor*{jesus} \citeyear{jesus}; \citeauthor*{pessev} \citeyear{pessev}; \citeauthor*{iaus266} \citeyear{iaus266}).  

Quite recently, we presented our own independent effort to provide models for interpreting integrated stellar cluster photometry.  Our model, MASSCLEAN (\textbf{MASS}ive \textbf{CL}uster \textbf{E}volution and \textbf{AN}alysis\footnote{\url{http://www.physics.uc.edu/\textasciitilde popescu/massclean/} \\ \texttt{MASSCLEAN} package is freely available unde GNU General Public License } -- \citeauthor*{masscleanpaper} \citeyear{masscleanpaper}), is rather unique in its design compared to other stellar cluster models and provides additional insight to interpreting presently available data on stellar clusters.  In this paper we first present in \S 2 our recently expanded data-set of 70 million Monte Carlo models, MASSCLEAN{\fontfamily{ptm}\selectfont \textit{colors}} (\citeauthor*{paper2} \citeyear{paper2}). This database of models shows the stochastic and non-Gaussian behavior of integrated stellar cluster colors and magnitudes as a function of cluster mass and age.  This, our first, newly completed, extended database of models, was selected to match the mass, age and metallicity range expected in the stellar clusters of the Large Magellanic Cloud (LMC).   In \S 3, we demonstrate the utility of this extensive database.  When used with a statistical inference code we have designed within MASSCLEAN called, MASSCLEAN{\fontfamily{ptm}\selectfont \textit{age}}, we use the database to work backwards from the integrated magnitudes and colors of real LMC clusters, to derive their stellar ages and masses.  We also discuss our new age results as compared to previous age determinations based on the same photometric data, to provide initial tests on the cluster results obtained using MASSCLEAN{\fontfamily{ptm}\selectfont \textit{age}}.  Final discussion of our results and conclusions are found in \S 4.

\section{The MASSCLEAN{\fontfamily{ptm}\selectfont \textit{colors}} Database -- Now Over 70 Million Monte Carlo Simulations} 

We have computed the integrated colors and magnitudes as a function of mass and age for a simple stellar cluster. Our results -- the newest version of the MASSCLEAN{\fontfamily{ptm}\selectfont \textit{colors}} database (\citeauthor*{paper2} \citeyear{paper2}) -- are based on over $70$ million Monte Carlo simulations
 using Padova 2008 (\citeauthor*{padova2008} \citeyear{padova2008}) and Geneva (\citeauthor*{geneva1} \citeyear{geneva1}) isochrones, and for two metalicities, $Z=0.019$ (solar) and $Z=.008$ (LMC). The simulation were done using Kroupa IMF (\citeauthor*{Kroupa2002} \citeyear{Kroupa2002}) with $0.1$ $M_{\Sun}$ and $120$ $M_{\Sun}$ mass limits. The age range of $[6.6,9.5]$ in $log(age/yr)$ was chosen (similar to \citeauthor*{masscleanpaper} \citeyear{masscleanpaper}; \citeauthor*{paper2} \citeyear{paper2}) to accomodate both Padova and Geneva models. The mass range is $200$-$100,000$ $M_{\Sun}$. An average of $5000$ clusters were simulated for each mass and age. 

In this work we will look closely at a subset of the MASSCLEAN{\fontfamily{ptm}\selectfont \textit{colors}} database.  We present integrated magnitudes and colors ($M_{V}$, $(U-B)_{0}$, $(B-V)_{0}$) representing 50 million Monte Carlo simulations using Padova 2008 stellar evolutionary models alone and for the single metallicity, $Z=.008$ ([Fe/H]=-0.6), selected to match the bulk properties of the LMC over most of the past few billion years (\citeauthor*{piatti2009} \citeyear{piatti2009}).

\subsection{The Distribution of Integrated Colors and Magnitudes as a Function of Mass and Age}

In \citeauthor*{paper2} \citeyear{paper2}, we analysed the influence of the stochastic fluctuations in the mass function of stellar clusters on their integrated colors as a function of cluster mass. Using 13 mass intervals ranging from 500--100,000 $M_{\Sun}$, we showed the mean value of the distribution of integrated colors varies with mass.  We further showed that the dispersion of integrated colors, measured by $1 \sigma$ (standard deviation), increases as the value of the cluster mass decreases. In this new work we will present not just the sigma range and mean, but the full distribution of integrated colors and magnitudes.  We are also using a much larger number of Monte Carlo simulations and cluster mass values in the 200--100,000 $M_{\Sun}$ interval.  This allows us to look closely at where the distribution of real cluster properties lie in the critical color-age and color-color diagrams.

\begin{figure*}[htp] 
\includegraphics[angle=270,width=16.0cm]{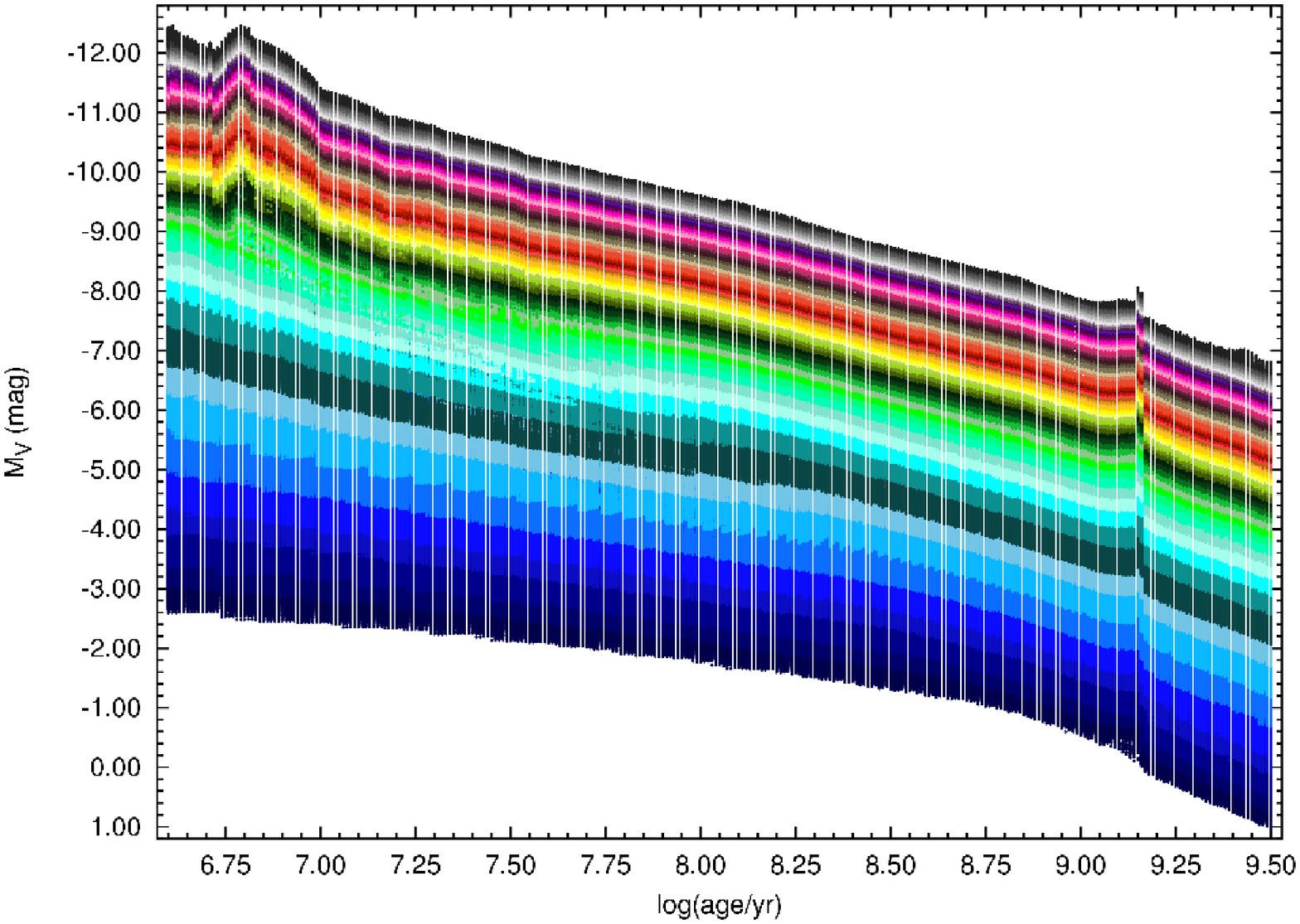}

\hspace{3.5 cm}
\includegraphics[angle=270,width=10.0cm, bb=165 82 445 755]{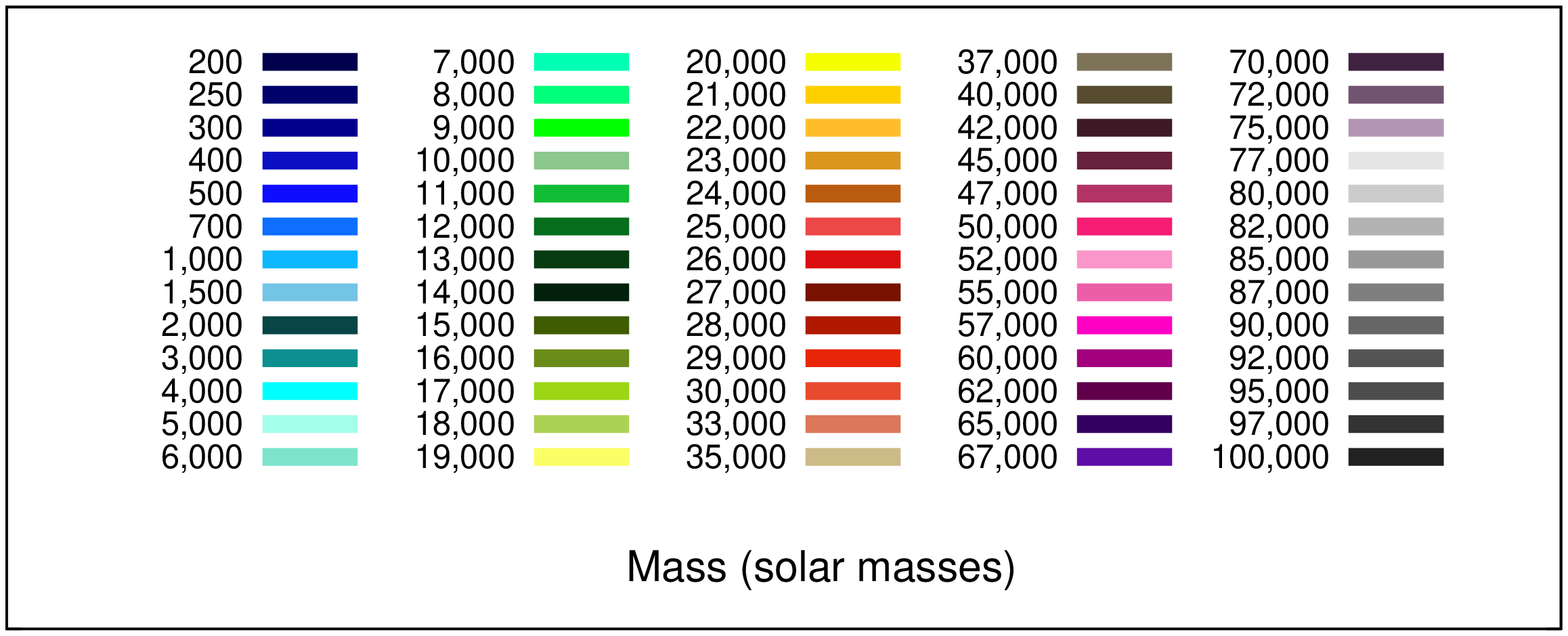}
\caption{\small The results from 50 million Monte Carlo simulations created with MASSCLEAN.  Models were generated with 65 different masses and evolved following Padova 2008 evolutionary models.  Higher mass clusters are plotted over preceded lower mass clusters. This figure gives the (partially correct) impression that $M_{V}$ monotonically increases with cluster mass for all ages. \normalsize}\label{fig:one2}
\end{figure*}

The distribution of integrated $M_{V}$ magnitudes as a function of mass and age is presented in the upper panel of Figure \ref{fig:one2}. The corresponding mass values are color-coded and presented in the bottom panel.  Sixty-five (65) different mass intervals have been explored in the MASSCLEAN{\fontfamily{ptm}\selectfont \textit{colors}} database as part of this demonstration.
  All 65 mass intervals are plotted on top of each other, making it difficult to see the individual distributions.  However, to first order, higher mass clusters have higher absolute magnitudes, $M_{V}$. 

\begin{figure*}[htp] 
\centering
\includegraphics[angle=270,width=16.0cm]{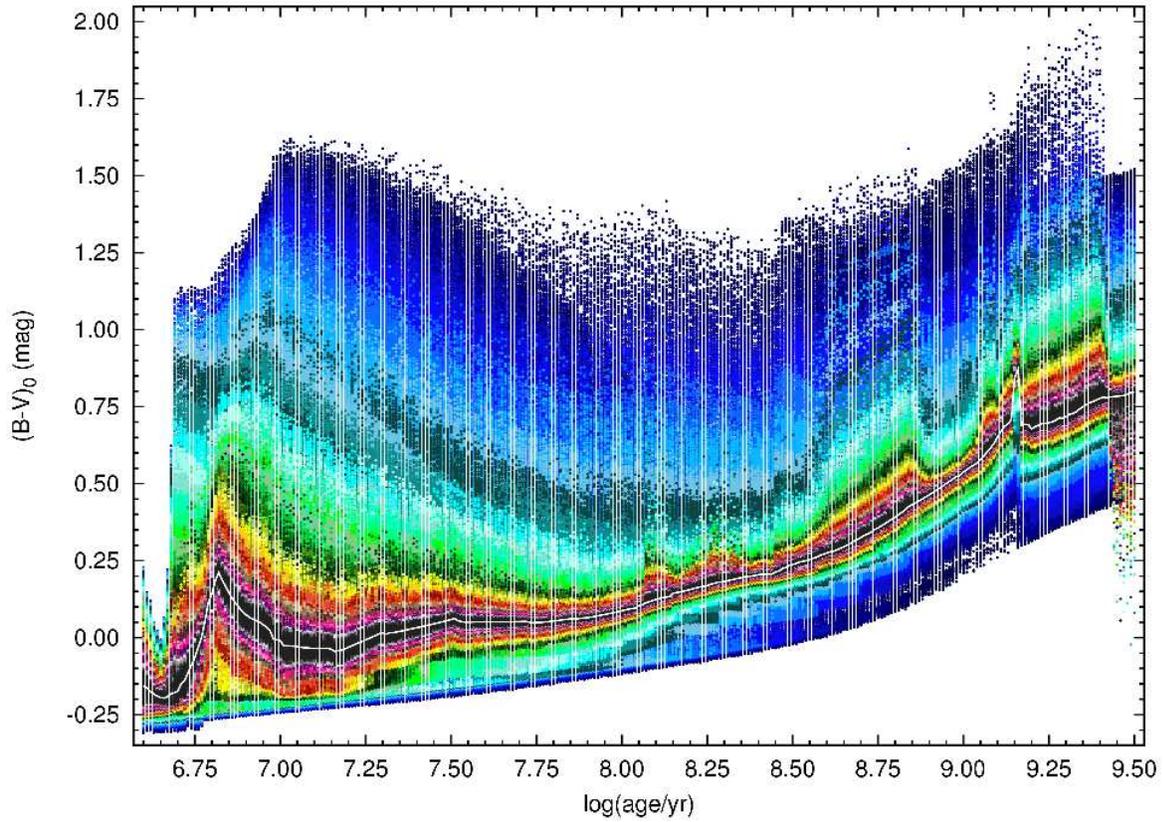}
\caption{\small The results from the same 50 million Monte Carlo simulations as shown in Figure \ref{fig:one2} and using the same color coding to represent cluster mass.  The range of $(B-V)_{0}$ colors, as a function of mass and age is given.  Again, higher mass clusters are plotted over preceded lower mass clusters, so covers the real distribution of expected colors as a function of mass and age.  A white line which lies at the color center, represents the color dependence with age computed in the infinite mass limit ($10^{6} M_{\Sun}$ in our simulations). This figure gives the (correct) impression that the range of cluster colors observed as a function of age, is greatly increased for lower mass clusters.\normalsize}\label{fig:two2}
\end{figure*}

\begin{figure*}[htp] 
\centering
\includegraphics[angle=270,width=16.0cm]{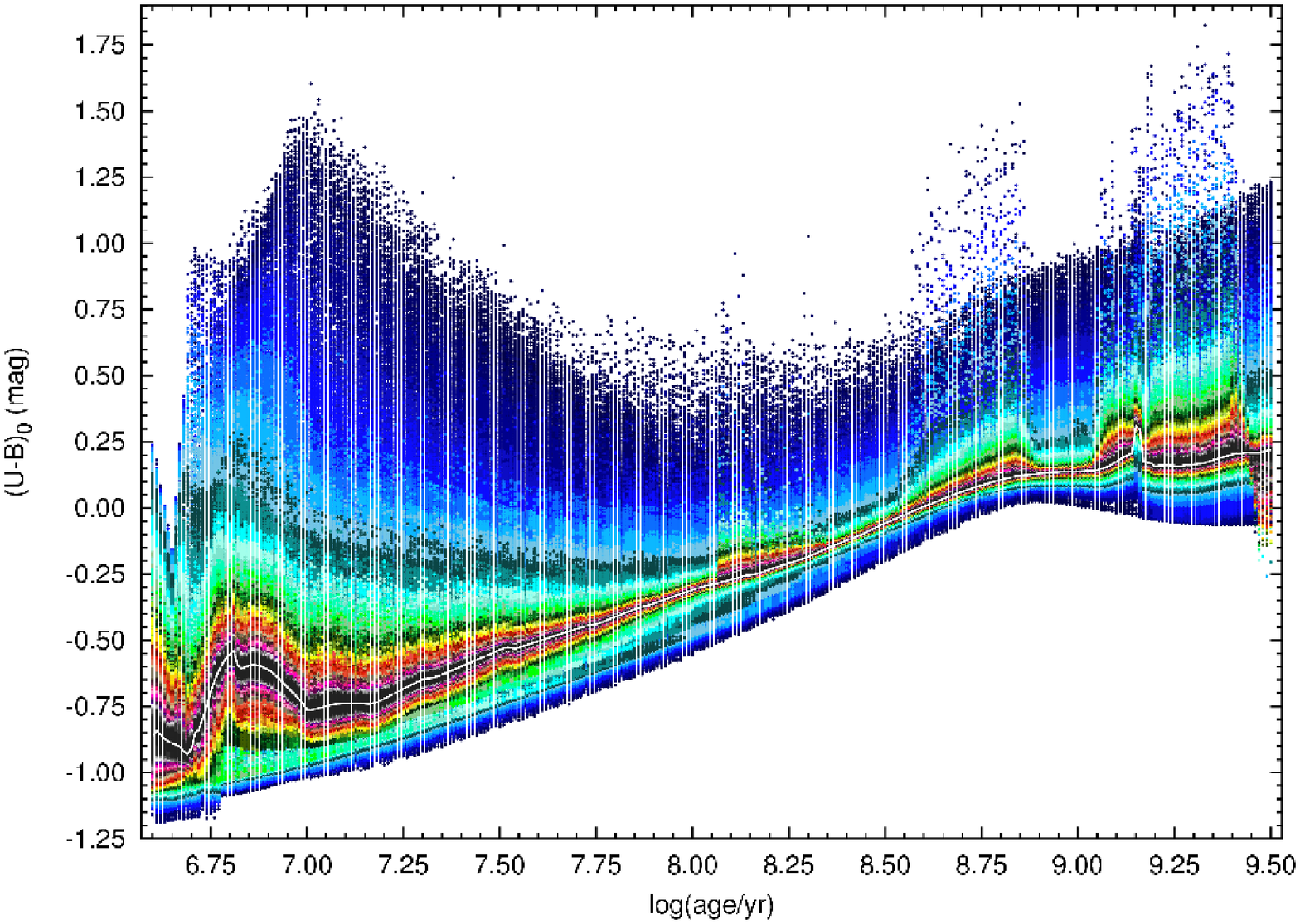}
\caption{\small The same as Figure \ 2, only this figure shows $(U-B)_{0}$ colors as a function of mass and age. \normalsize}\label{fig:three2}
\end{figure*}

In Figures \ref{fig:two2} and \ref{fig:three2} we present the distribution of $(B-V)_{0}$ and $(U-B)_{0}$ integrated colors as a function of mass and age. The mass values are represented by the same colors as shown in the bottom panel of Figure \ref{fig:one2}. The variation of integrated colors computed in the infinite mass limit ($10^{6}$ $M_{\Sun}$ in our simulations) and consistent with standard simple stellar population (SSP) models based on Padova 2008 (\citeauthor*{padova2008} \citeyear{padova2008}), is represented by the solid white line in both figures.  As was shown in \citeauthor*{paper2} \citeyear{paper2}, the dispersion in integrated colors grows larger for lower mass clusters, both in the blue and red side. As the cluster mass increases, the dispersion decreases and the integrated colors approach the infinite mass limit shown in white.  While it is valuable to see the range of masses over-layed in a single plot, it is difficult to see the exact shape of the color distributions underlying each mass interval.

\subsection{The Shape of the Distribution of Integrated Colors and Magnitudes}

\begin{figure*}[htp] 
\centering

\includegraphics[angle=270,width=5.25cm]{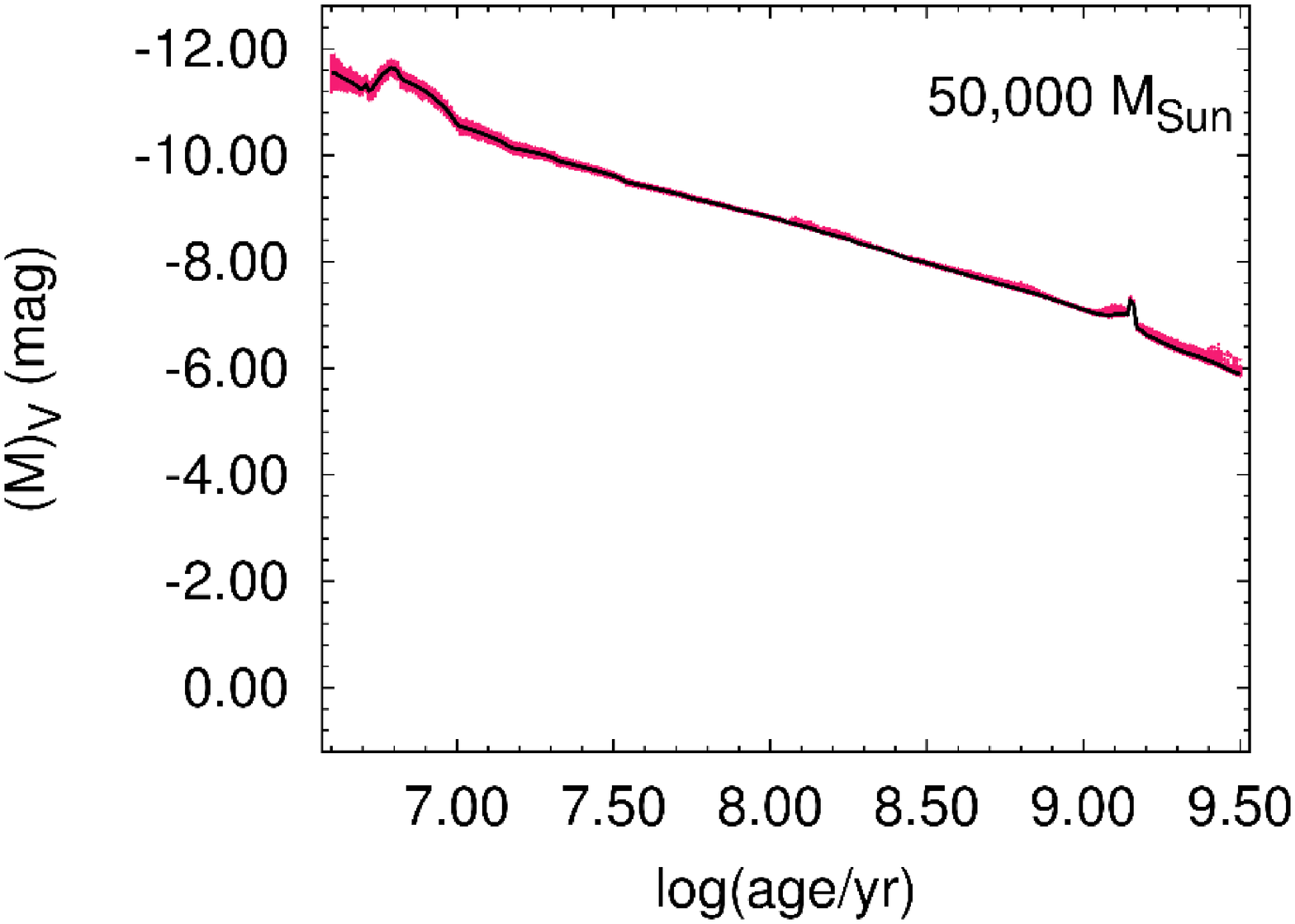}
\includegraphics[angle=270,width=5.25cm]{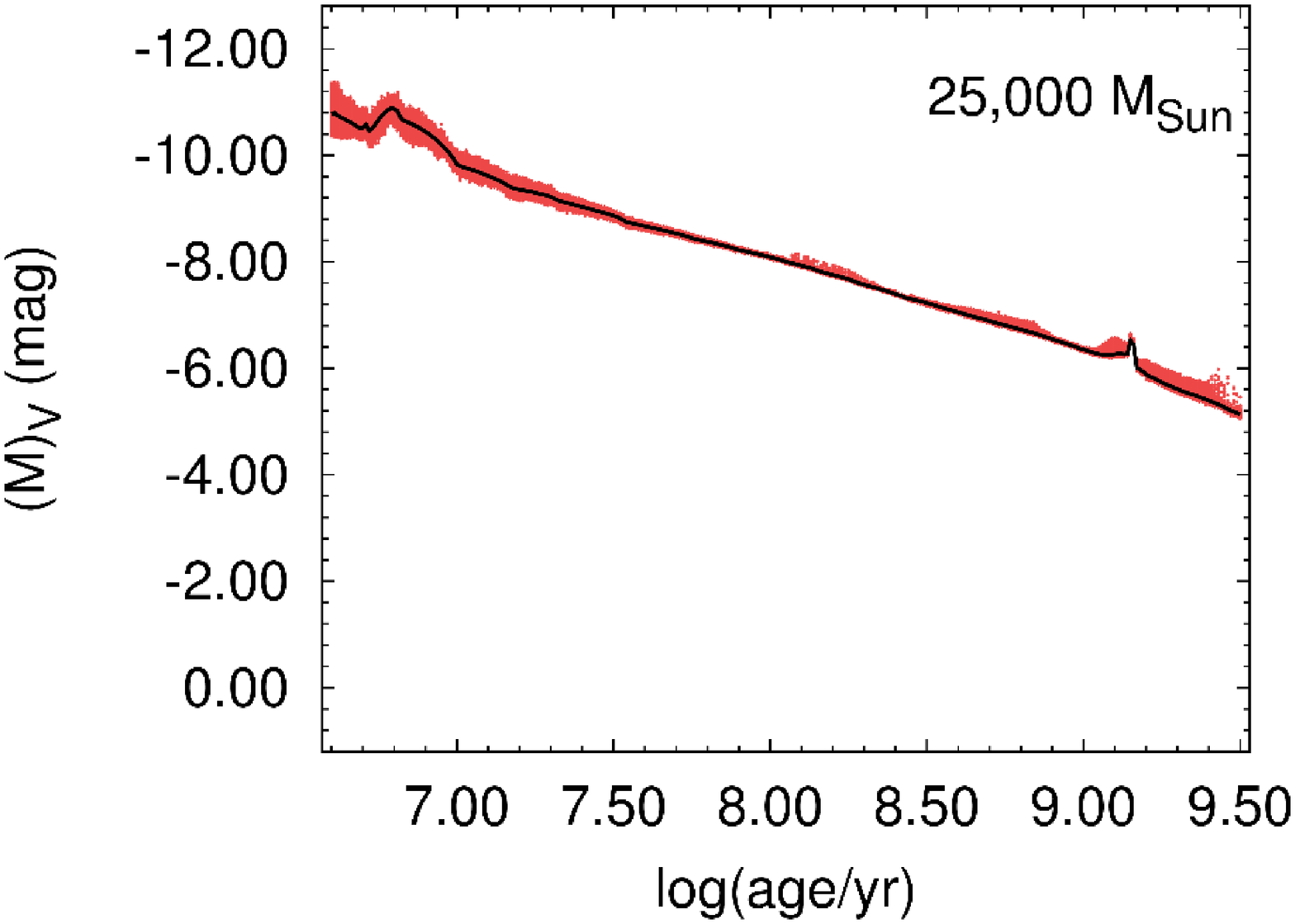}
\includegraphics[angle=270,width=5.25cm]{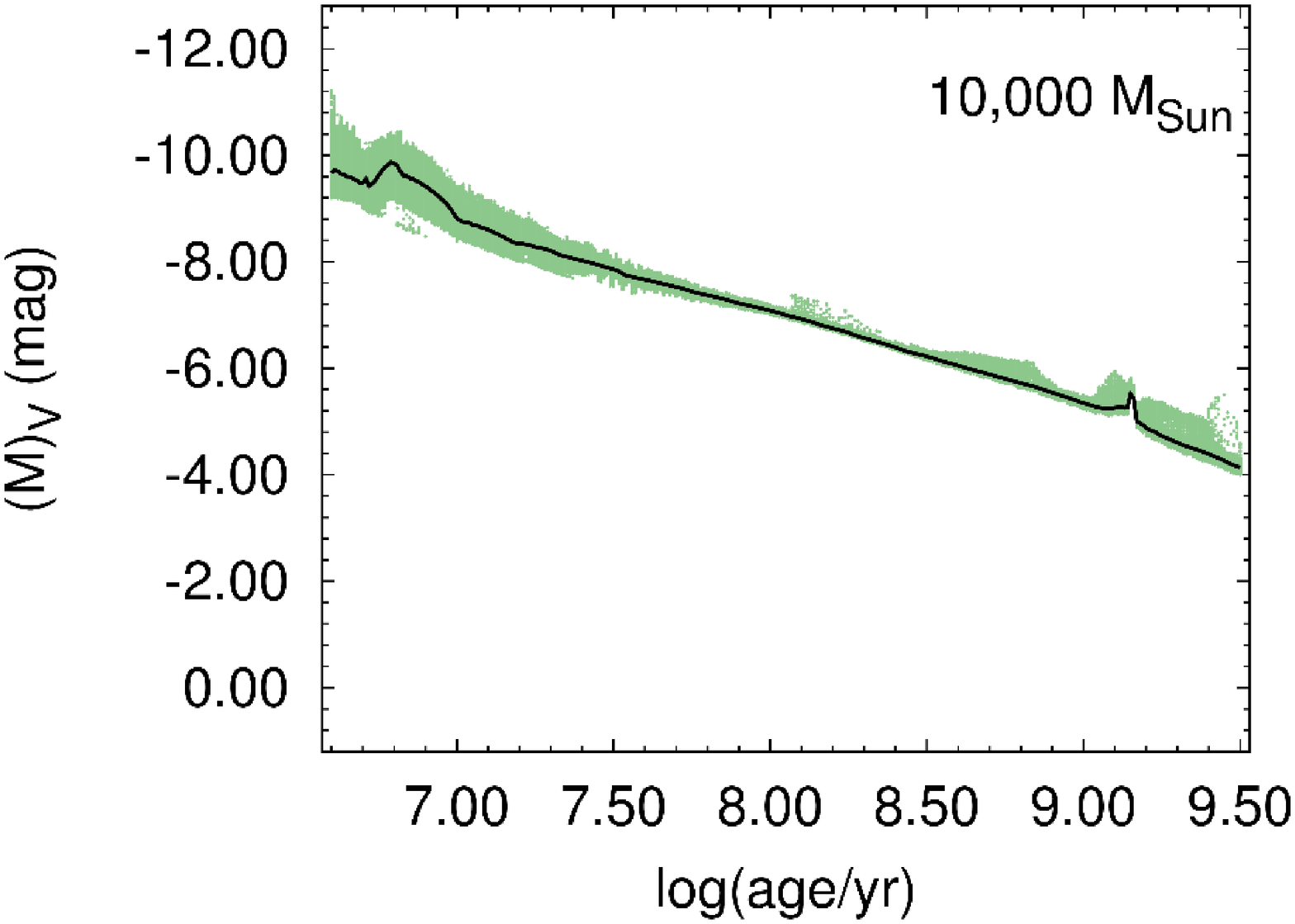}

\includegraphics[angle=270,width=5.25cm]{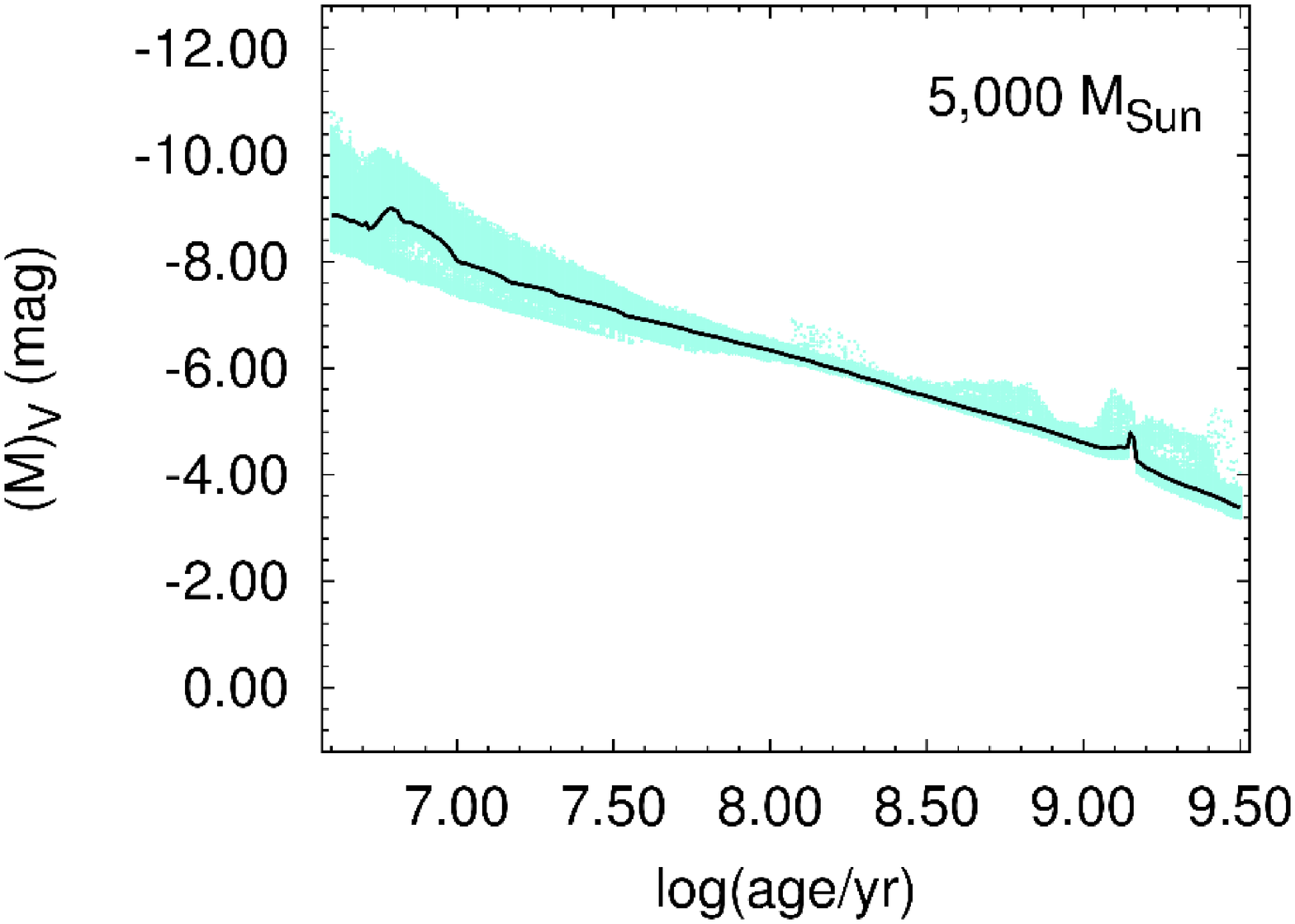}
\includegraphics[angle=270,width=5.25cm]{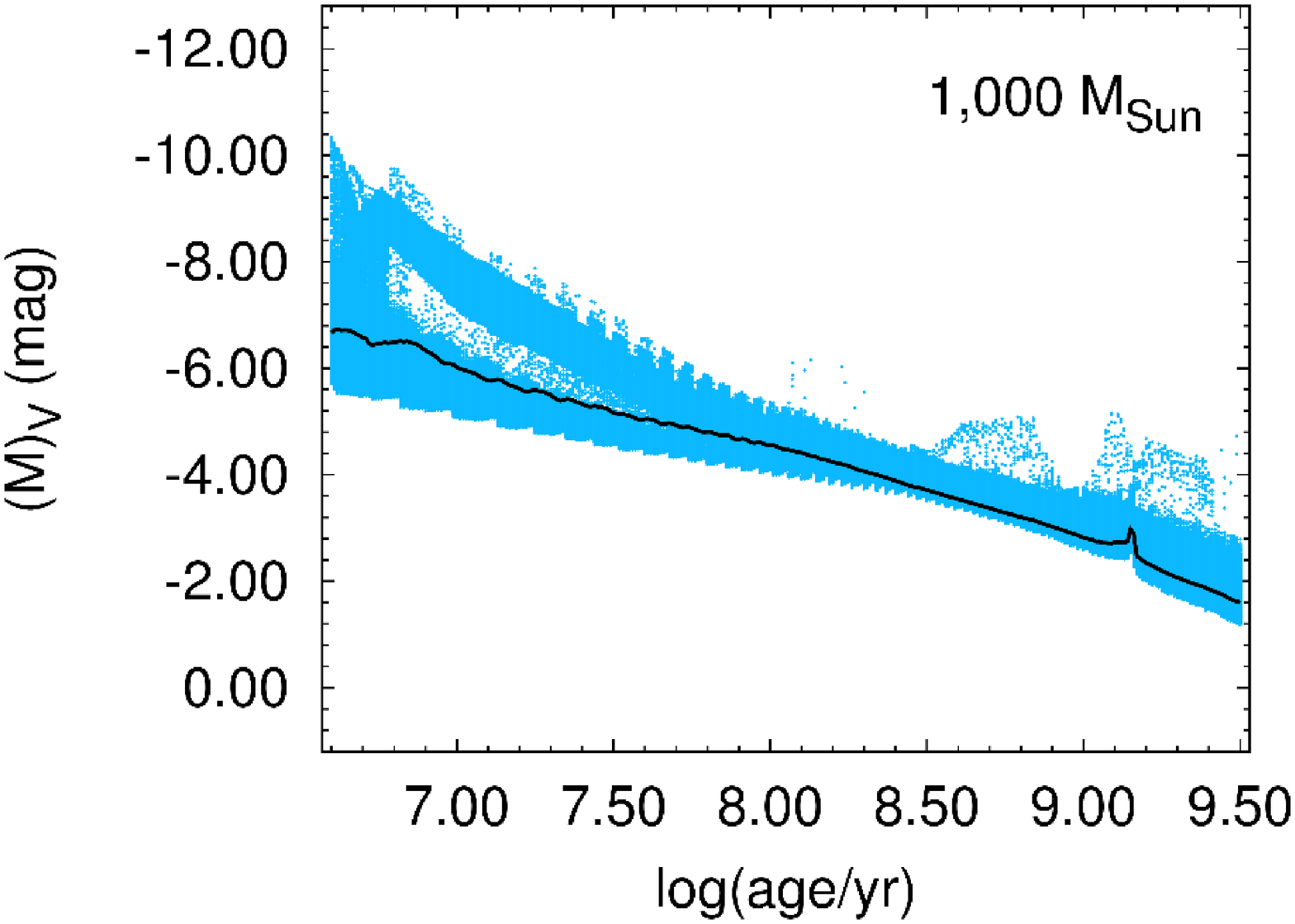}
\includegraphics[angle=270,width=5.25cm]{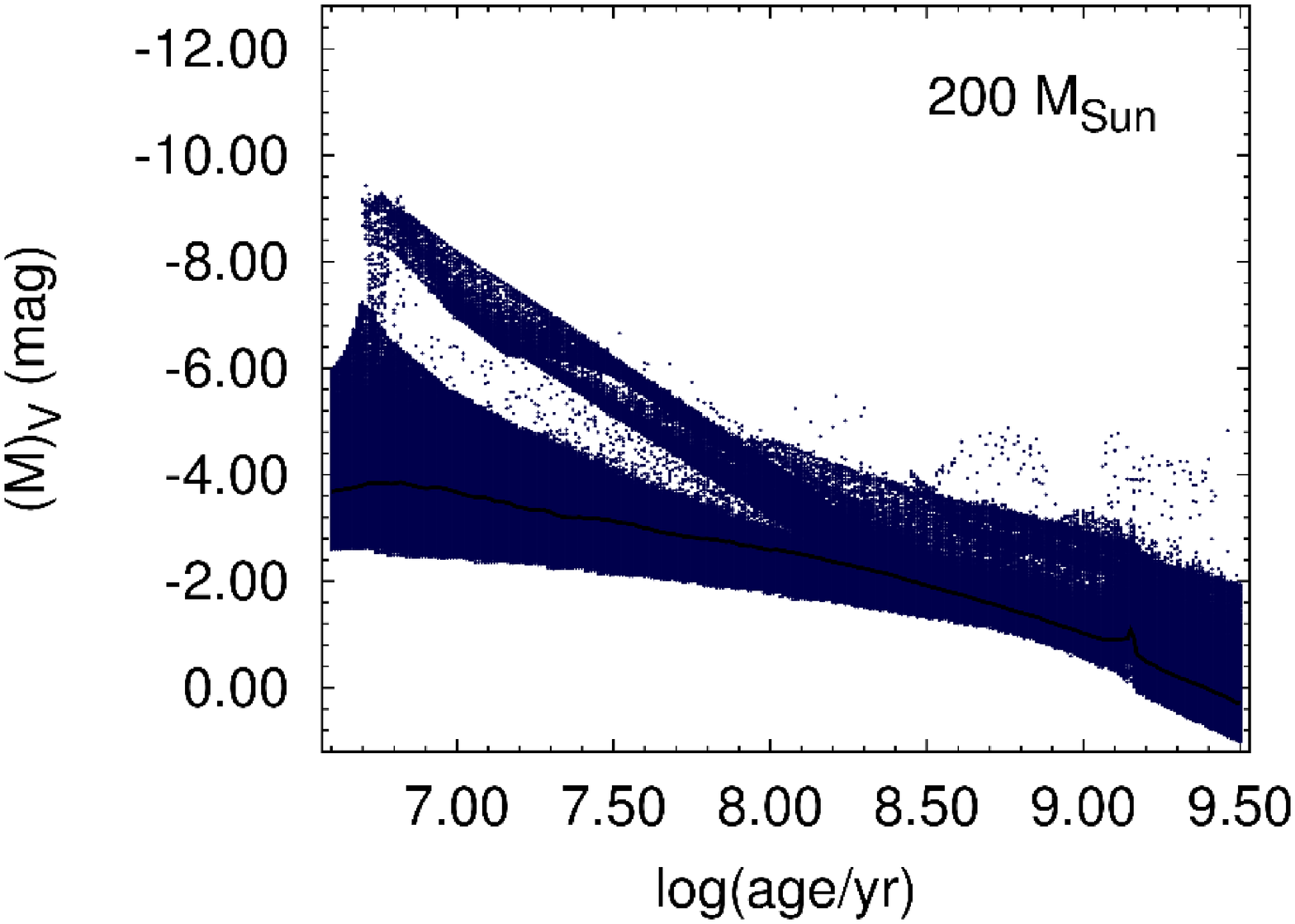}

\caption{\small Beginning in the upper left panel, the black line shows the mean value of the distribution of integrated $M_{V}$ colors for a 50,000 $M_{\Sun}$ cluster.  The red points are MASSCLEAN, Monte Carlo derived observed magnitudes for a 50,000 $M_{\Sun}$ cluster, as a function of age.  Over one million models are used to derive the red points in this figure.  The distribution seen about the mean value is relatively small.  In the next panel to the right, is given the magnitudes, as a function of age, for Monte Carlo derived cluster with mass, 25,000 $M_{\Sun}$.  The black line is the mean value of the distribution of integrated $M_{V}$ colors for a 25,000 $M_{\Sun}$ cluster. In the following 4 subfigures, we provide lower and lower mass simulated clusters, 10,000 (upper right), then 5,000 (lower left), 1,000 (lower center) and finally 200 $M_{\Sun}$ (lower right), all the while showing the mean value of the distribution of integrated $M_{V}$ colors for the corresponding clusters mass.  For clusters with masses of only a few thousand solar mass, the observed range becomes extremely large, and can even mimic clusters of lower or even considerably higher total mass.  \normalsize}\label{fig:four2}
\end{figure*}
\begin{figure*}[htp] 
\centering

\includegraphics[angle=270,width=5.25cm]{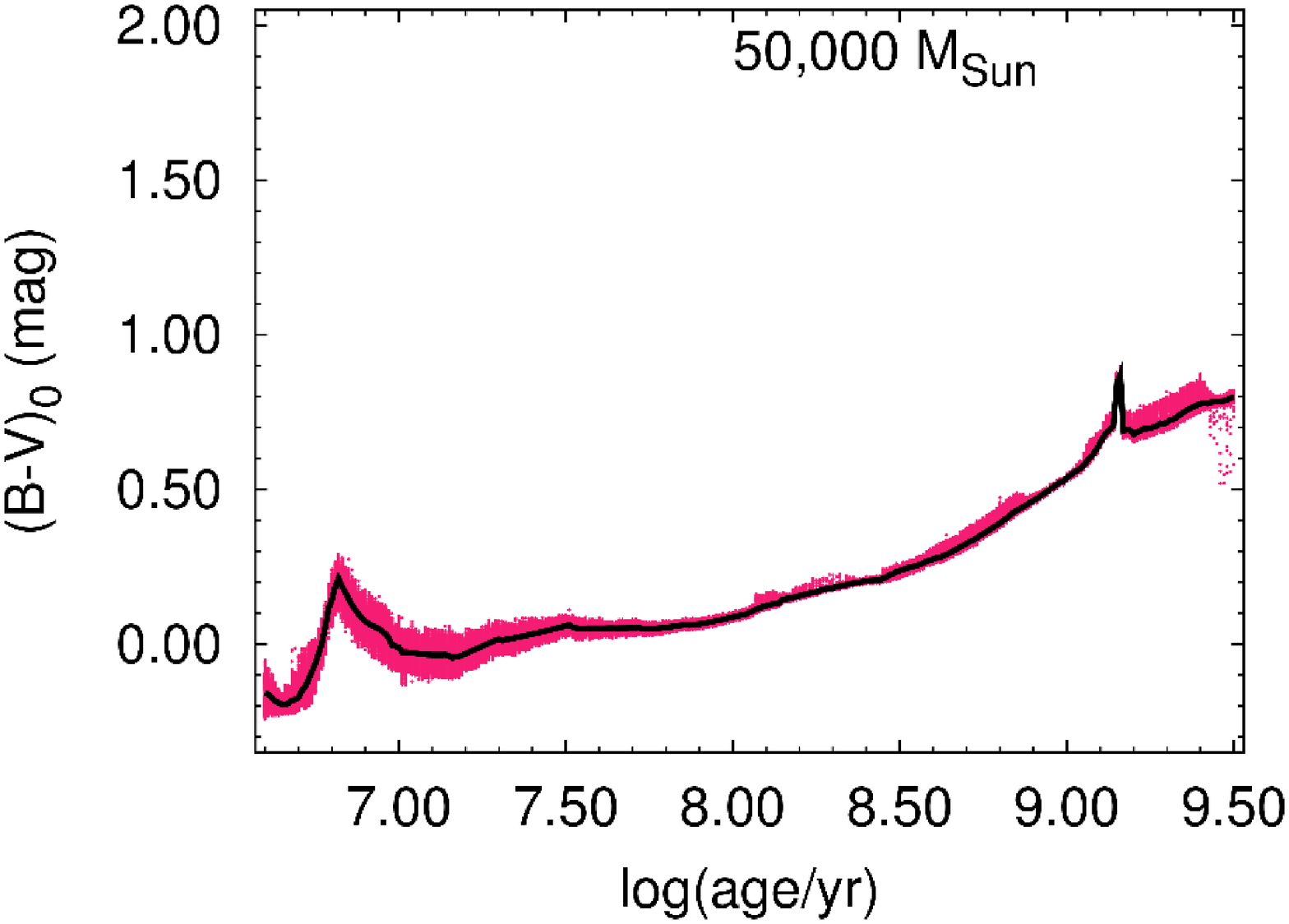}
\includegraphics[angle=270,width=5.25cm]{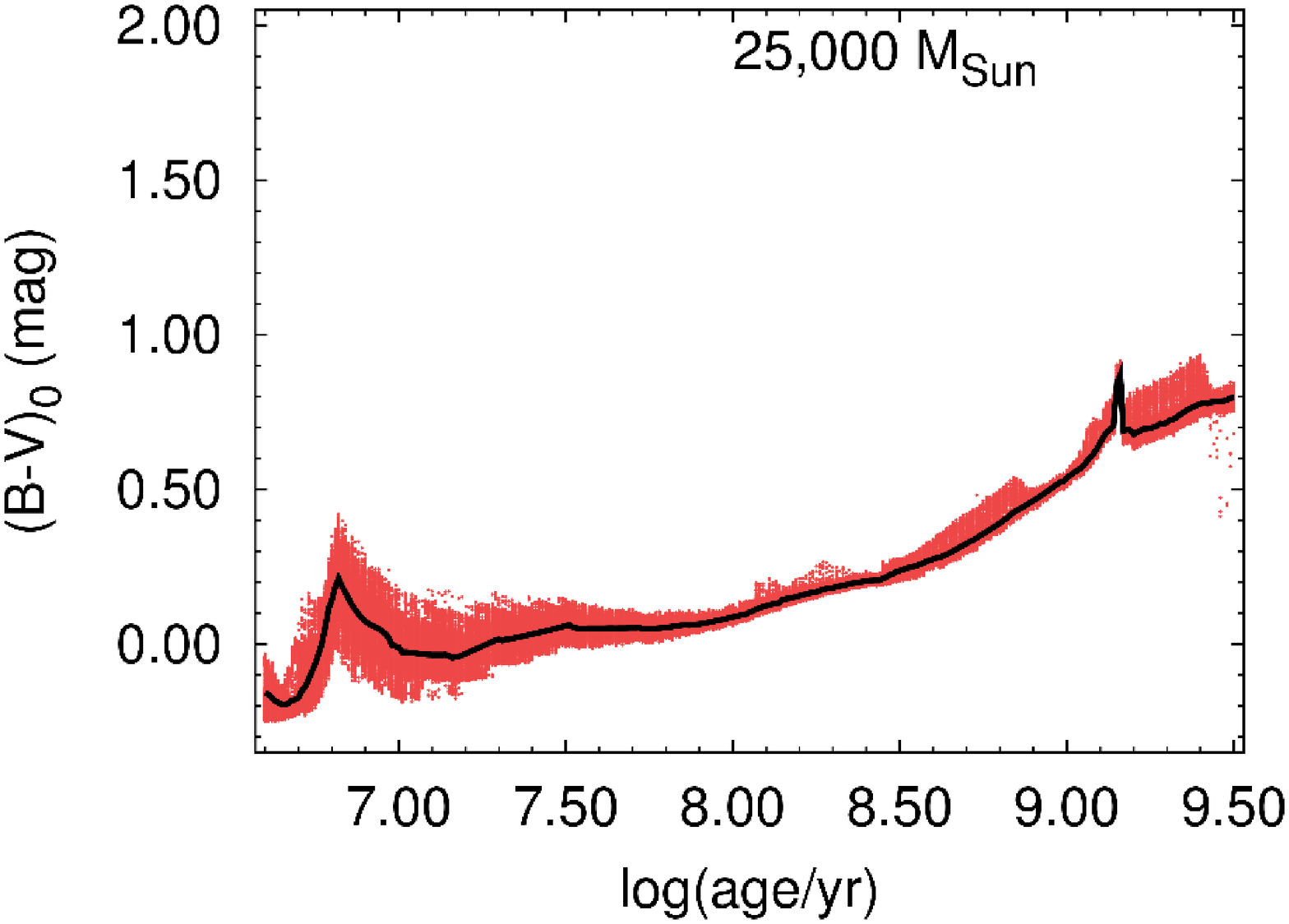}
\includegraphics[angle=270,width=5.25cm]{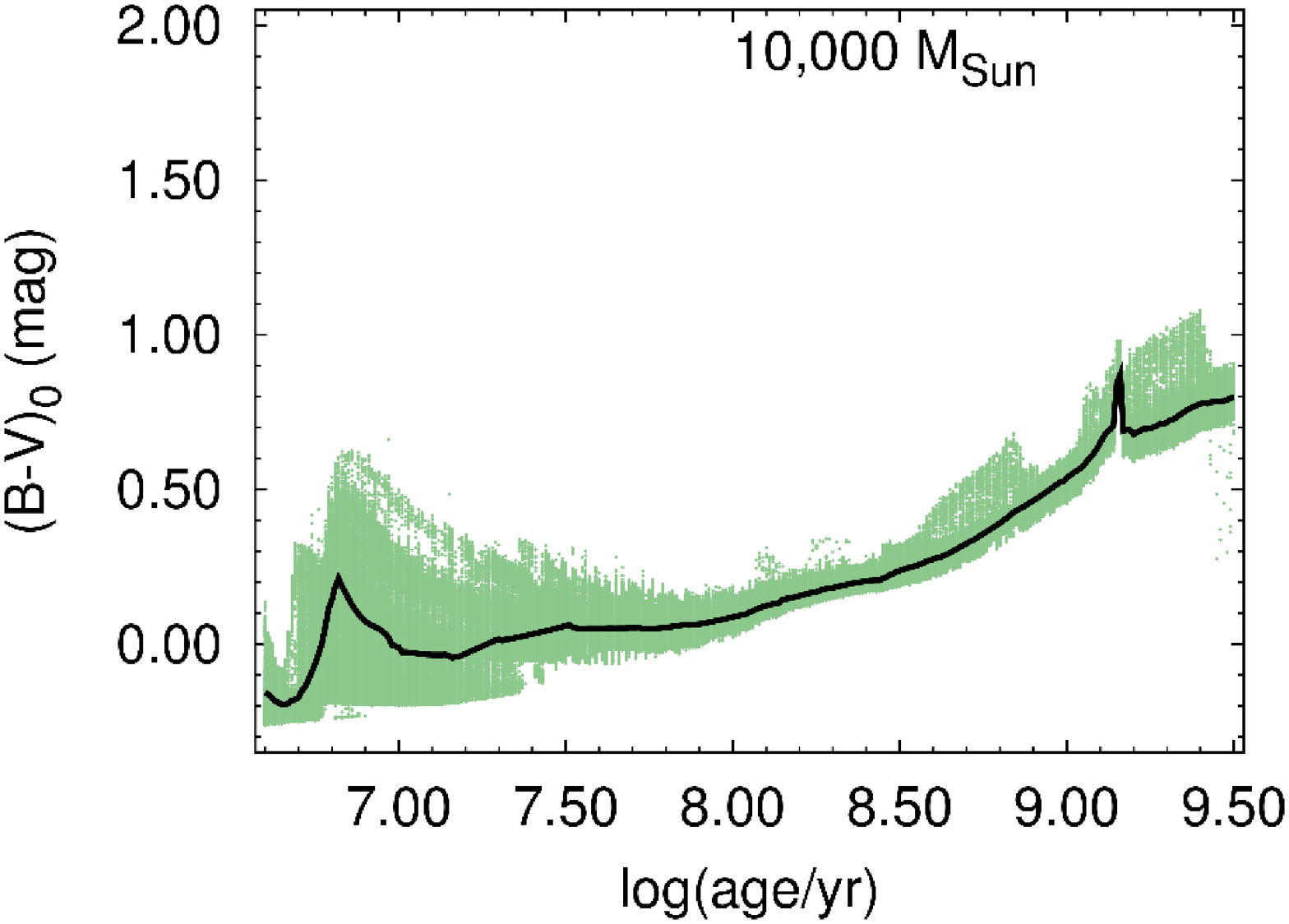}

\includegraphics[angle=270,width=5.25cm]{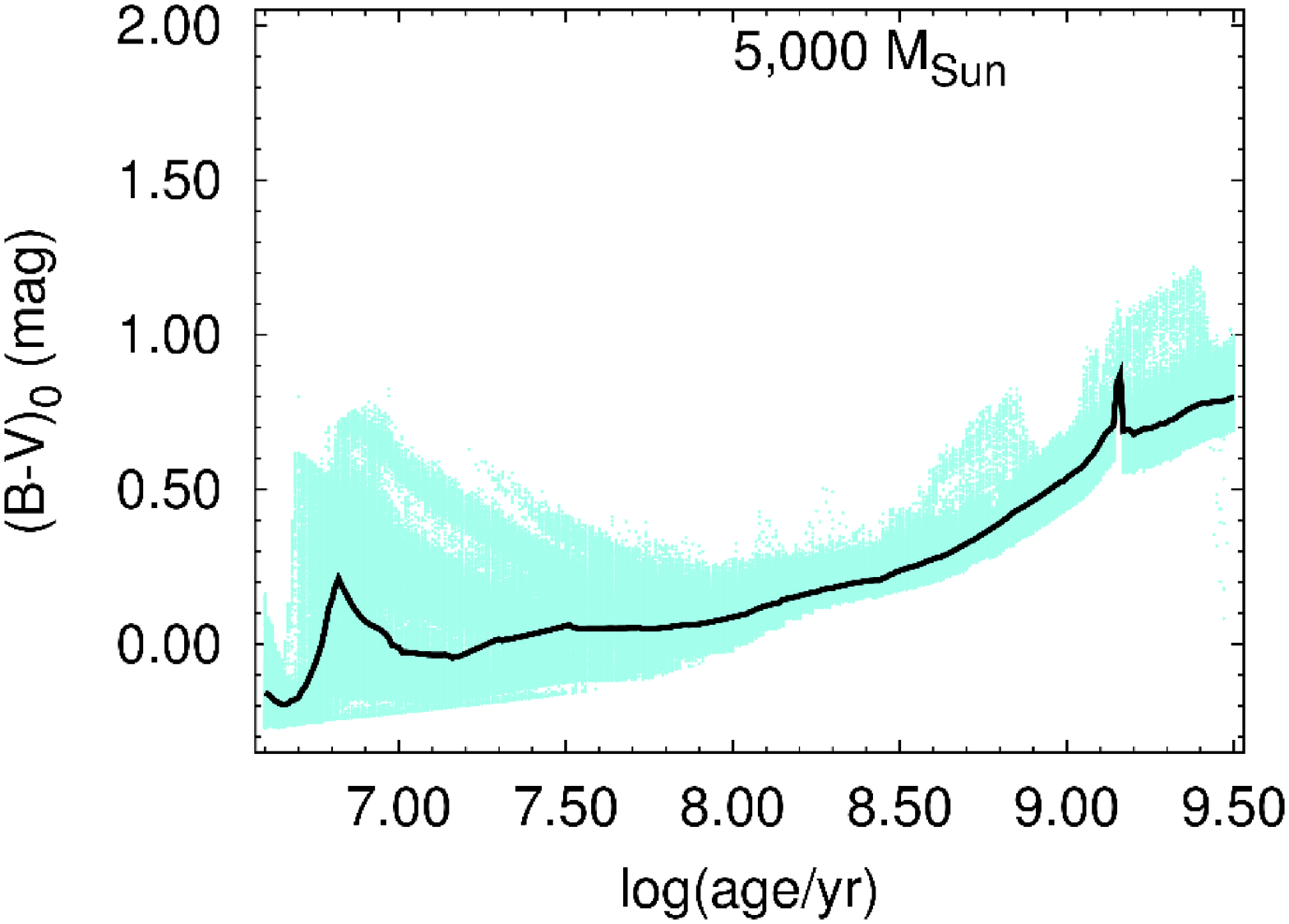}
\includegraphics[angle=270,width=5.25cm]{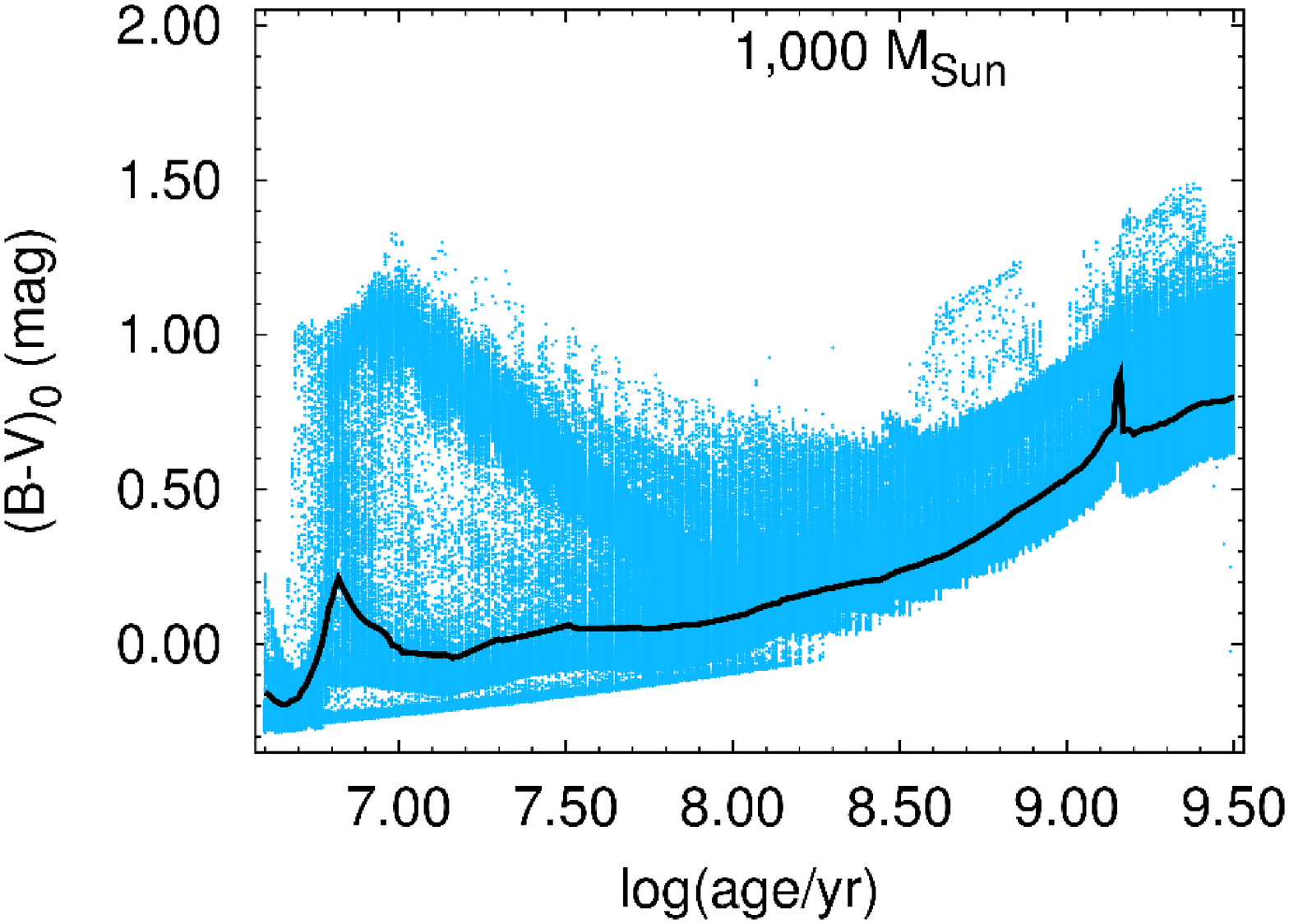}
\includegraphics[angle=270,width=5.25cm]{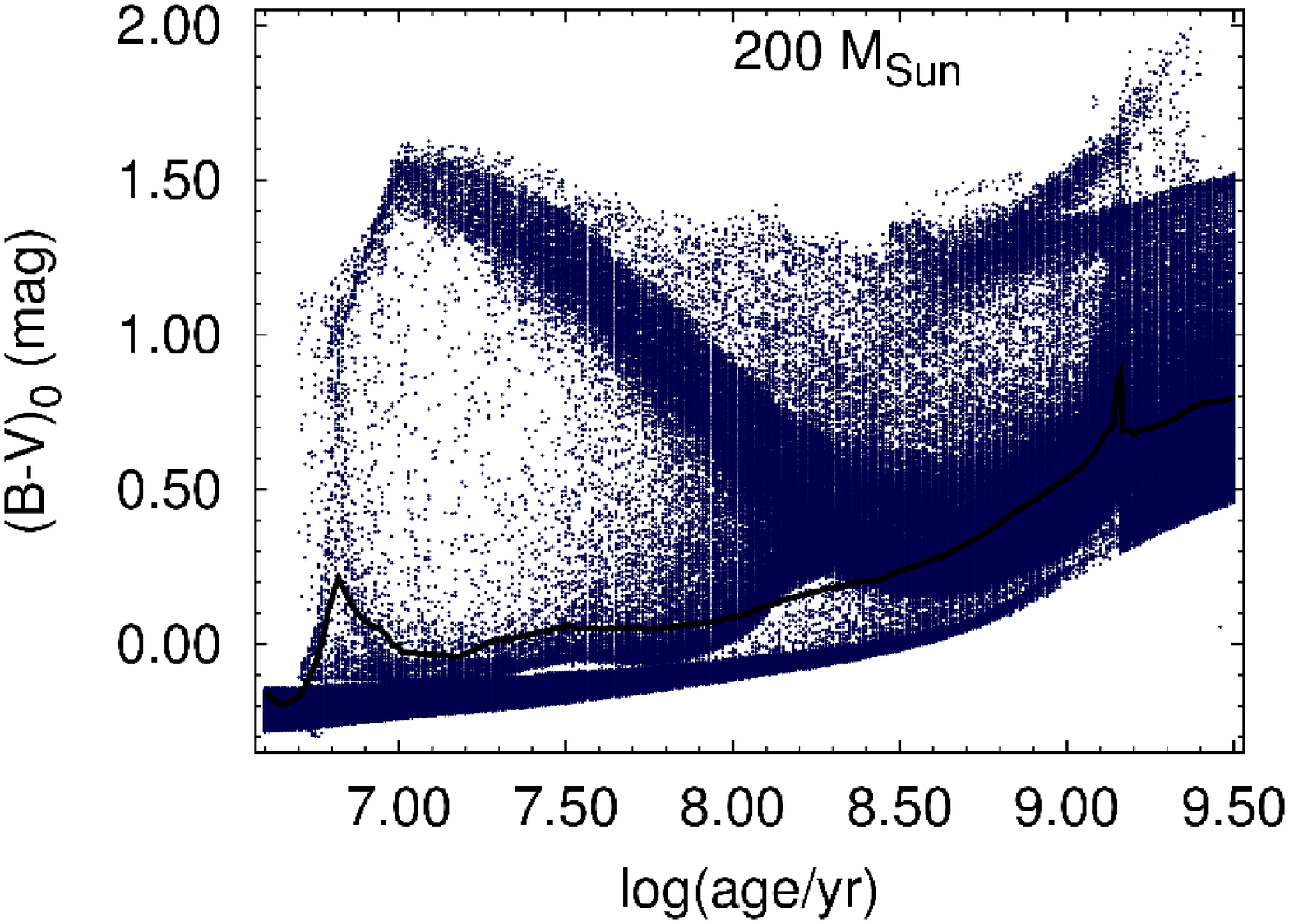}

\caption{\small Like with Figure \ref{fig:four2}, we have selected six mass examples (50,000, 25,000, 10,000, 5,000, 1000, and 200 $M_{\Sun}$) to show the observed distribution of $(B-V)_{0}$ colors as a function of age, compared to the infinite mass limit (in the black line, the same for all six plots).  For the higher mass clusters, 50,000 and 25,000 $M_{\Sun}$, the range of colors is small and stays reasonably close to the values computed in the infinite mass limit.  However, in the remaining panels one sees that the observed values from our simulations can demonstrate rather extreme colors, bluer, but more so very red colors at times, for increasingly lower mass clusters.\normalsize}\label{fig:five2}
\end{figure*}
\begin{figure*}[htp] 
\centering

\includegraphics[angle=270,width=5.25cm]{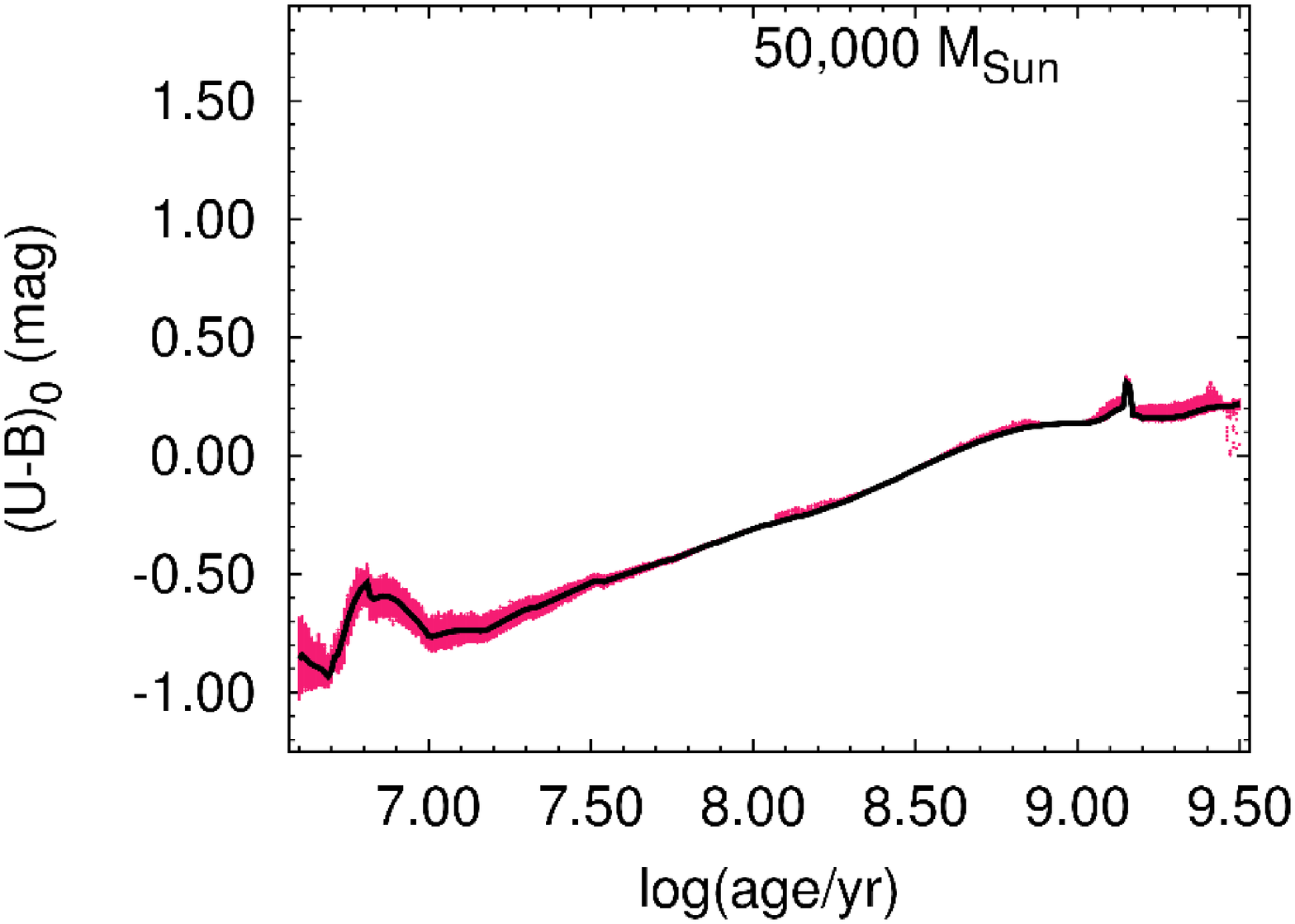}
\includegraphics[angle=270,width=5.25cm]{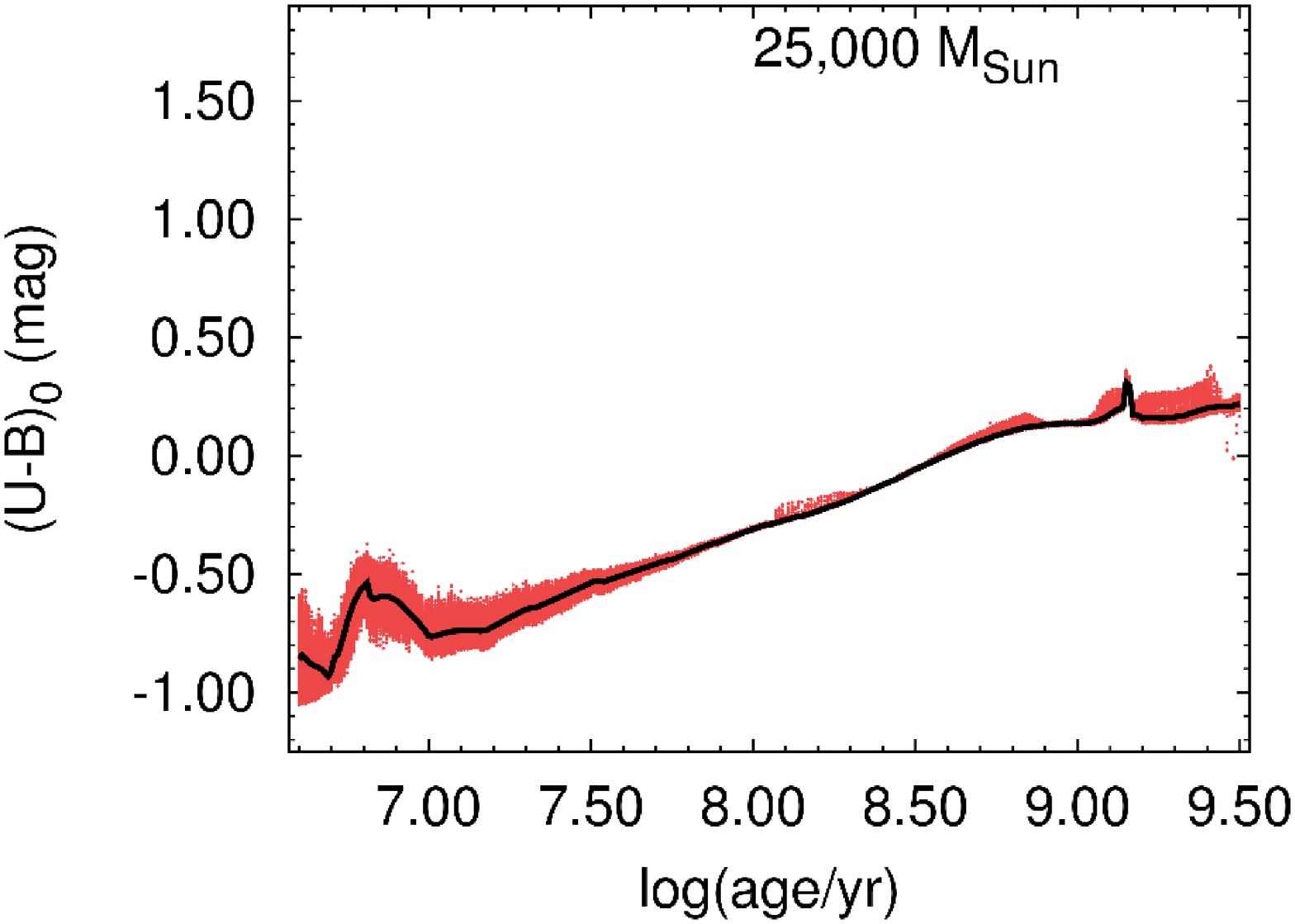}
\includegraphics[angle=270,width=5.25cm]{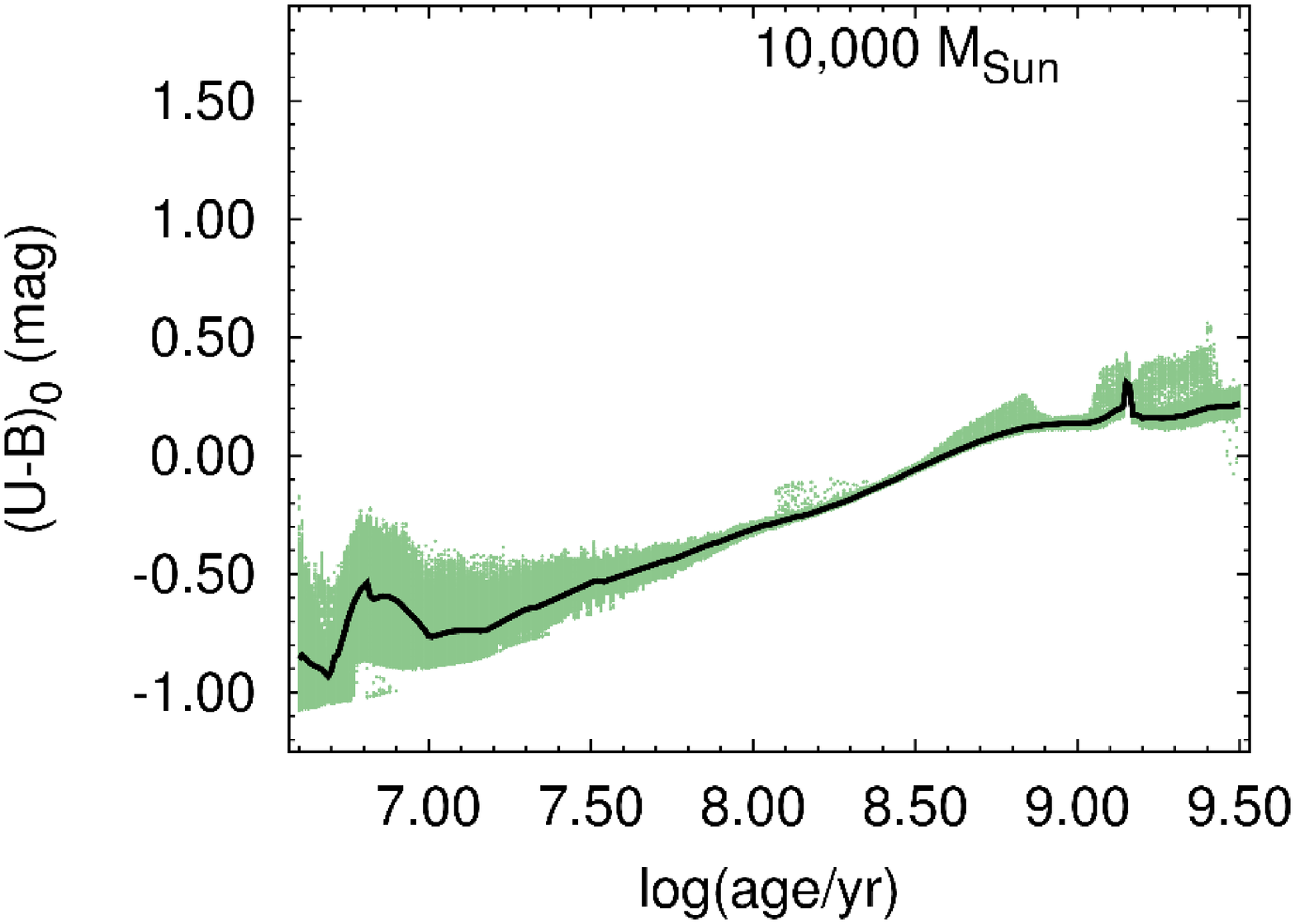}

\includegraphics[angle=270,width=5.25cm]{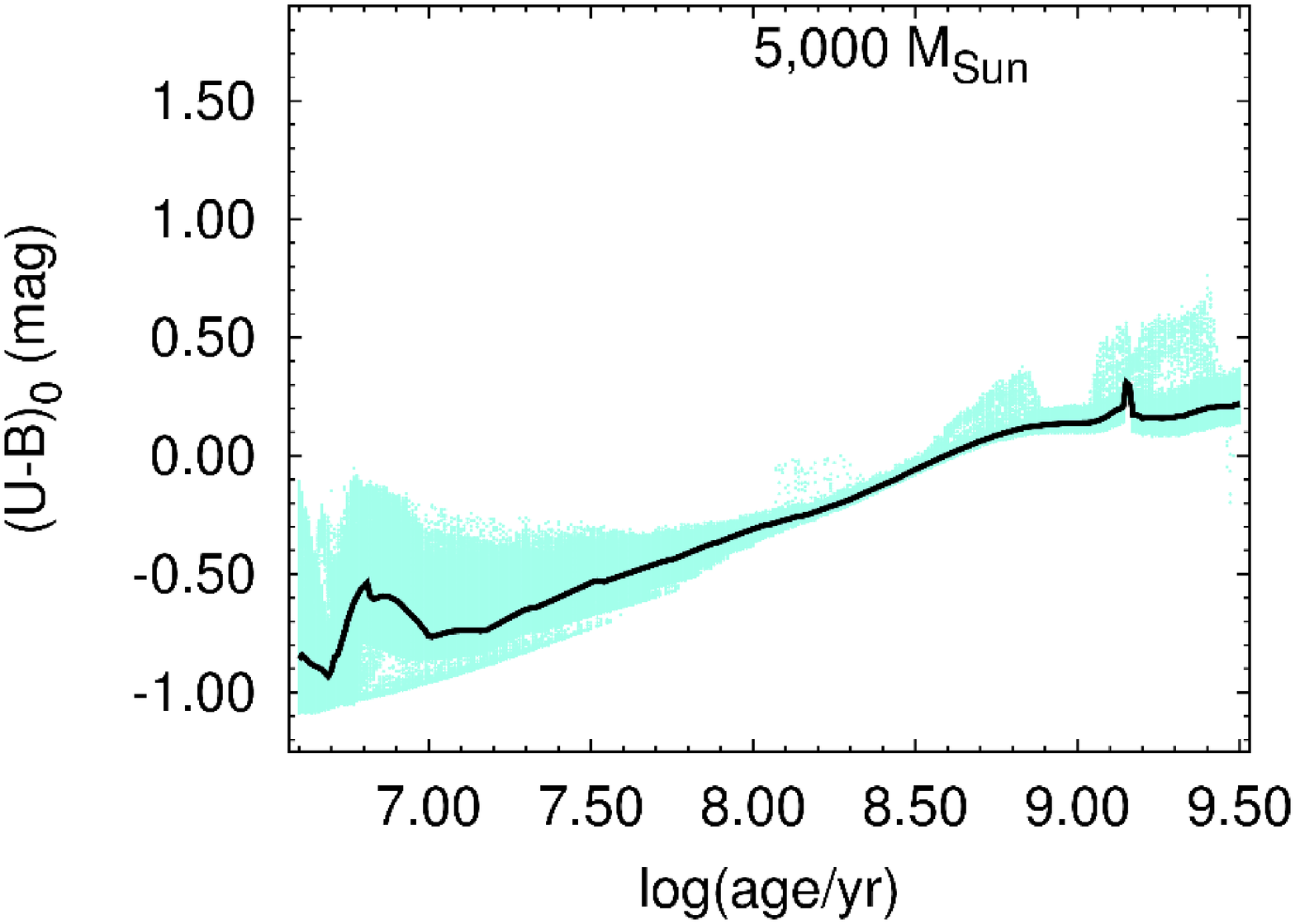}
\includegraphics[angle=270,width=5.25cm]{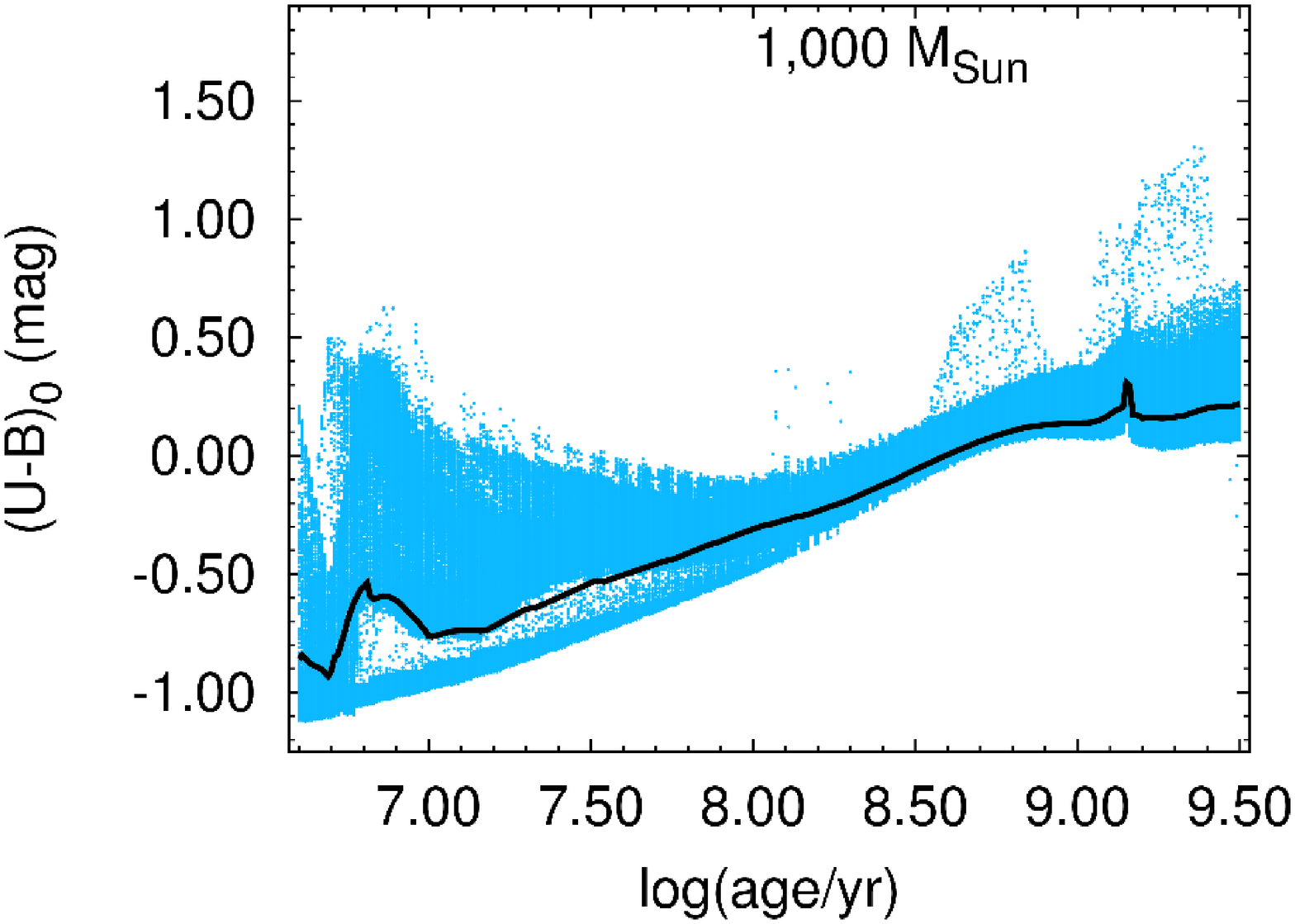}
\includegraphics[angle=270,width=5.25cm]{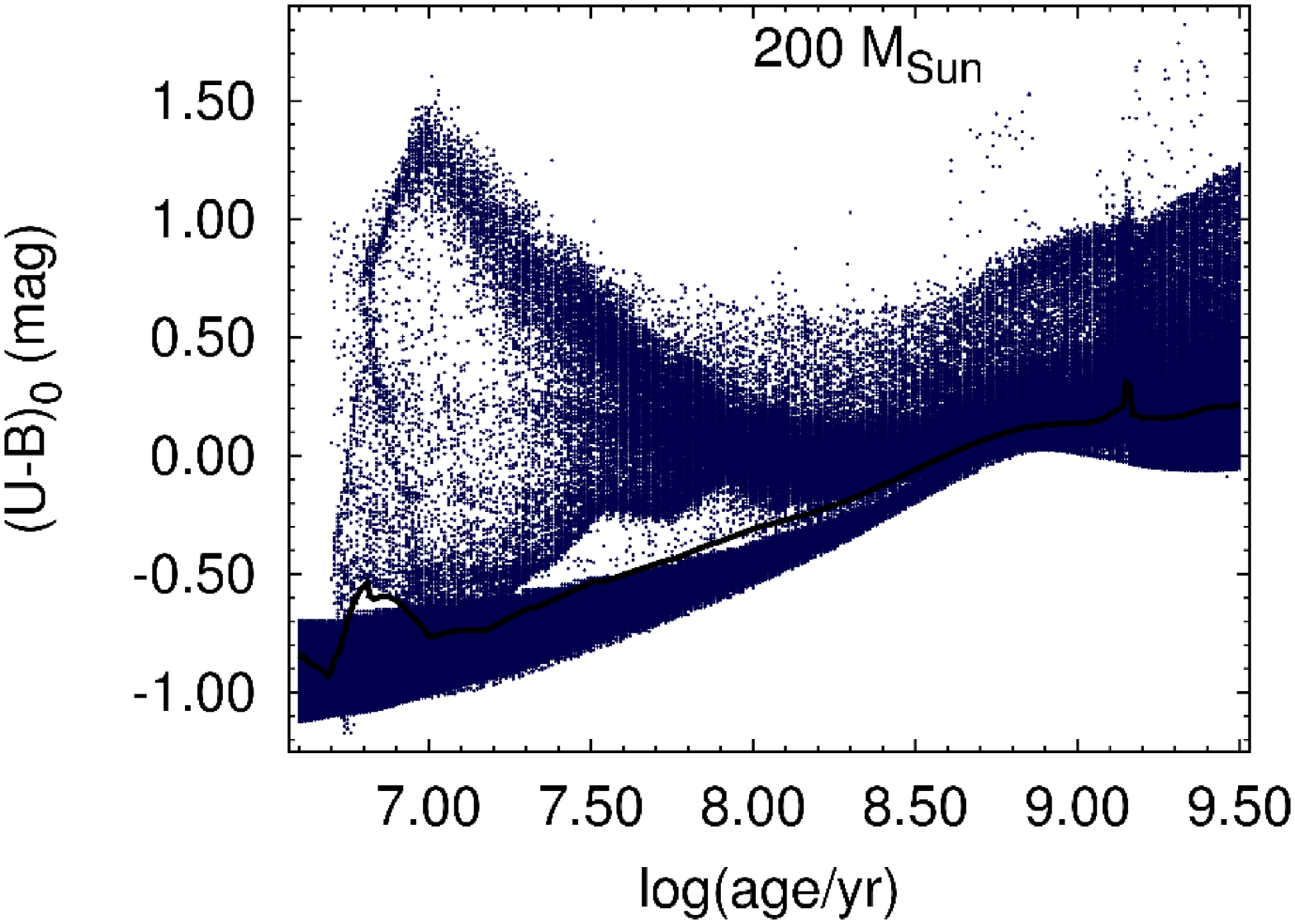}

\caption{\small This is the same as for Figure \ref{fig:five2}, only here we show the variation with mass and age of $(U-B)_{0}$ colors, as derived through Monte Carlo simulations using MASSCLEAN. \normalsize}\label{fig:six2}
\end{figure*}

We have selected a few single values of mass for independent plots in Figures \ref{fig:four2}--\ref{fig:six2}. 
The distribution of integrated $M_{V}$ magnitudes is presented in Figure \ref{fig:four2}, for clusters mass of 200, 1,000, 5,000, 10,000, 25,000, and 50,000 $M_{\Sun}$, respectively. The colors are the same as the ones used in Figure \ref{fig:one2}. What is also given in a black line is the mean value of the distribution of integrated $M_{V}$ magnitudes for that mass (\citeauthor*{paper2} \citeyear{paper2}). Several very critical effects are immediately apparent. The distribution (range) of observed integrated magnitudes for low mass clusters gets very large with lower mass clusters.  This was already presented in \citeauthor*{paper2} \citeyear{paper2}. However, the shape of that dispersion was not presented like we have done here.   Here we see in Fig.\ \ref{fig:four2}, particularly with 200 and 1,000 solar mass clusters, the distribution becomes {\sl bimodal} for the lowest mass clusters, especially for young ages (log t $\le$ 8.0).  A cluster with no more than 200 solar masses can, at times, have an absolute magnitude typical of a cluster $25-100$ times more massive and at the same age.  Even if one can derive an accurate age for their cluster, one can not rely on the absolute magnitude alone to derive mass unambiguously.  

In Figures \ref{fig:five2} and \ref{fig:six2} we present the distribution of integrated colors $(B-V)_{0}$ and $(U-B)_{0}$ as a function of age for the same select values of cluster mass, 200, 1,000, 5,000, 10,000, 25,000, and 50,000 $M_{\Sun}$, respectively. The integrated colors as a function of age, computed in the infinite mass limit ($10^{6} M_{\Sun}$ in our simulations) is represented by the solid black line.  Here the structure in the values observed start out closely aligned with the classical SSP codes predictions (infinite mass limit) for the high mass clusters, but quickly fans out to a rather broad distribution below 10,000 $M_{\Sun}$.  When one gets below 1,000 $M_{\Sun}$, the color range is extreme, both in the red and blue colors predicted by our models.  Moreover, the distribution shows a clear bimodal distribution (\citeauthor*{paper2} \citeyear{paper2}; \citeauthor*{lancon2002} \citeyear{lancon2002}; \citeauthor*{cervino2004} \citeyear{cervino2004}; \citeauthor*{fagiolini2007} \citeyear{fagiolini2007}). In \citeauthor*{paper2} \citeyear{paper2}, we discussed the increased dispersion in colors as the cluster mass decreased, though we did not investigate masses as low as shown here. However, even clusters with masses of 1,000 to as much as 5,000 $M_{\Sun}$ show an extraordinary range in colors, away from the predicted SSP single expectation value with age.

\subsection{The Distribution of Integrated Colors in the Color-Color Diagram}

\begin{figure*}[htp] 
\centering
\includegraphics[angle=0,width=16.0cm]{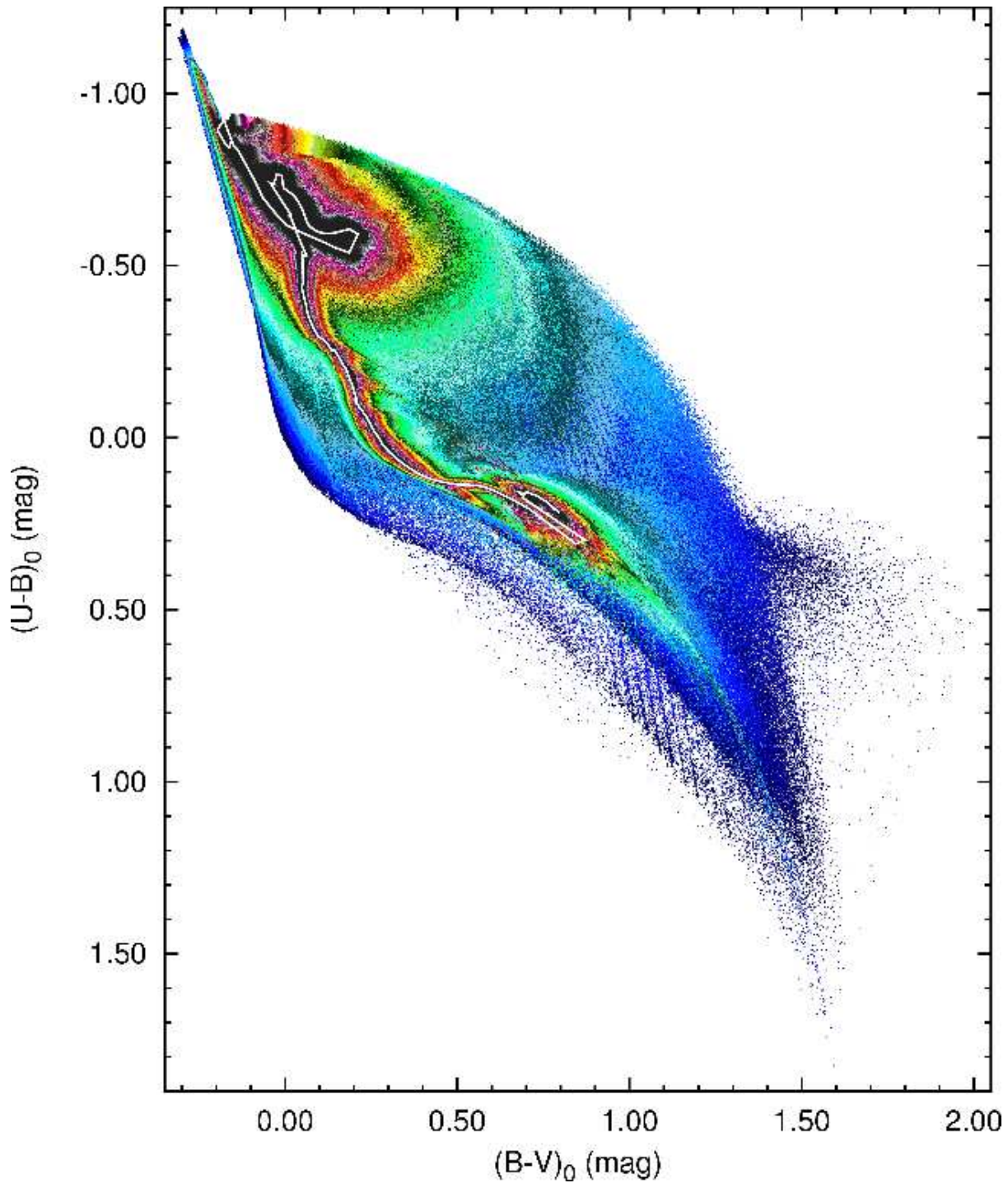}
\caption{\small The result of 50 million Monte Carlo simulations, color coded by mass as described in Figure \ref{fig:one2}.  Here we see the full range of color-color magnitudes predicted as a function of mass.  Clusters more massive than 50,000 $M_{\Sun}$ (in the pink) typically adhere closely to the predicted values for the single solution Padova-based SSP predictions, computed in the infinite mass limit.  However, clusters with masses below 10,000 $M_{\Sun}$ (the dark green just beyond the yellow) can show extreme color ranges.   \normalsize}\label{fig:seven2}
\end{figure*}

All the data from Figures \ref{fig:two2} and \ref{fig:three2} are again presented in Figure \ref{fig:seven2} as a $(U-B)_{0}$ vs. $(B-V)_{0}$ color-color diagram. And, once again, the variation of integrated colors computed in the infinite mass limit (as predicted by classical SSP codes) is represented by the solid white line. The 65 masses given are color-coded similarly to Figure \ref{fig:one2}.  Provided the mass of the cluster is greater than 50,000 $M_{\Sun}$ (in the pink), the predicted properties of our Monte Carlo generated clusters stays reasonably close to the integrated colors computed in the infinite mass limit.  For clusters with masses less than 10,000 $M_{\Sun}$ (the dark green just beyond the yellow), the range of possible colors becomes quite large, overlapping heavily with many different aged clusters and causing enormous degeneracy problems.  

The {\it classical} method for cluster age determination using integrated colors is based mainly on the $(U-B)_{0}$ vs. $(B-V)_{0}$ color-color diagram. The large dispersion presented in Figure \ref{fig:seven2} shows the limitation of this method when only the infinite mass limit is used. To better view what is happening in the complex Figure \ref{fig:seven2}, we show another view of this data in Figure \ref{fig:eight2}. Here, we apply an entirely new color scheme to represent our Monte Carlo simulations.  In Figure \ref{fig:eight2} we plot the color-color diagram for 5 ages ($log(age/yr)$= 7.00, 7.50, 8.00, 8.50, 9.00) and all values of mass between 200--100,000 $M_{\Sun}$.  Clusters of all masses are shown now color coded based on age.  Combining this with what we have learned from Figure \ref{fig:seven2}, Figure \ref{fig:eight2} shows us that even very young low mass clusters can mimic the UBV colors of more massive old clusters.  The extreme tail of colors belongs solely to intermediate and low mass clusters.   It is easy to see that very young clusters ($log(age/yr)$= 7.00, 7.50) could be mistaken for 1 billion years old ($log(age/yr)$= 9.00) clusters, based on the single SSP relationship in a classical UBV diagram.

The $(U-B)_{0}$ versus $(B-V)_{0}$ color-color diagram is a very important tool in age determination of stellar clusters. The high level of dispersion shown in Figures \ref{fig:seven2} and \ref{fig:eight2} demonstrates the limitations of this method when only the infinite mass limit is used.
  The degeneracy seems to be so extreme, one might think all is lost.  But in fact, the purpose of our extensive database was not just to show the range and degeneracy of the colors with age and mass (and here we have only limited ourselves to a single IMF, chemistry and evolutionary model), but to see if by producing such a database, we might provide a reasonable grid of properties from which to work back and derive cluster ages and masses.

\subsection{Mass and Age Are Degenerate Quantities in UBV Colors}
\begin{figure*}[htp] 
\centering
\includegraphics[angle=0,width=16.0cm]{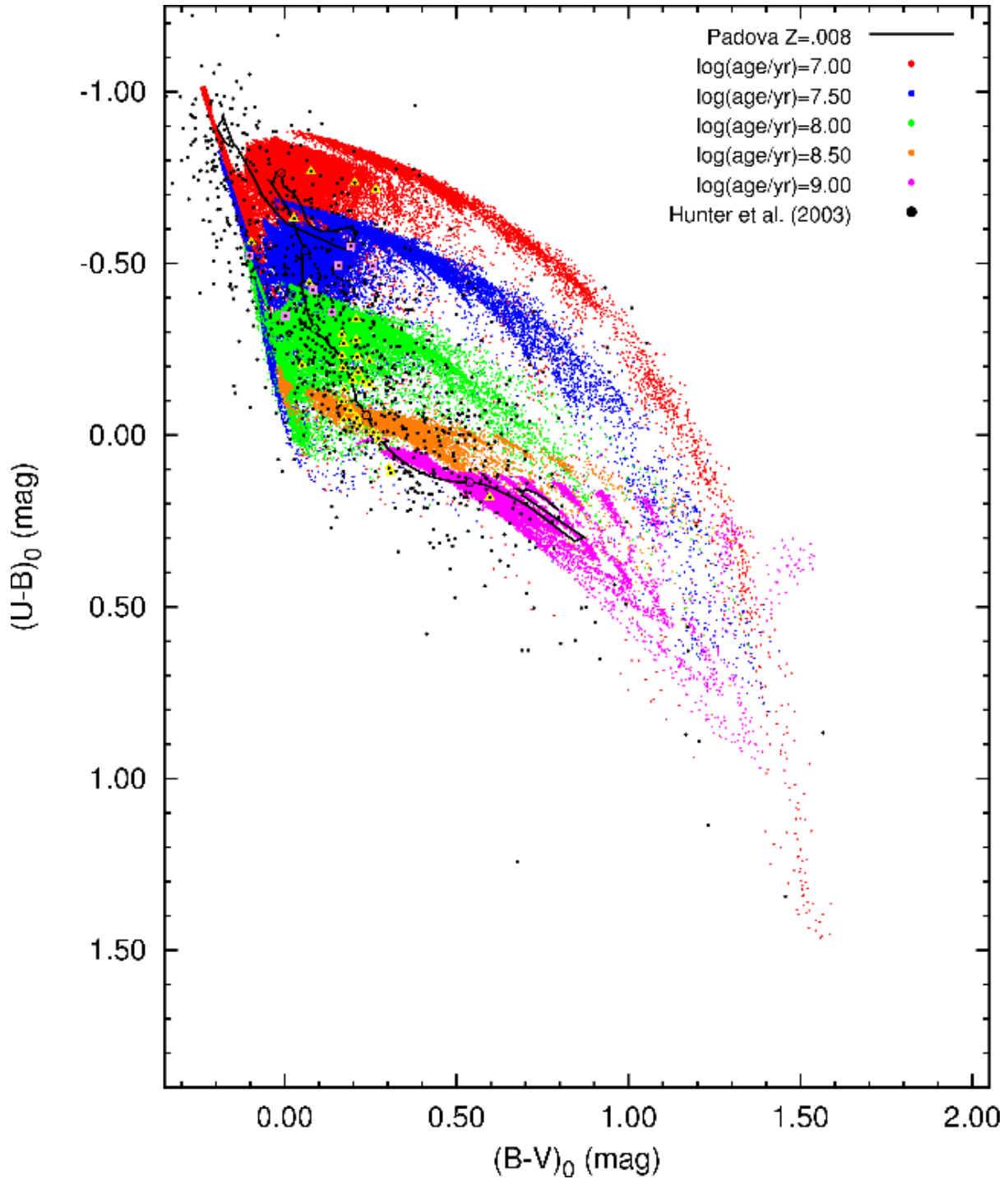}
\caption{\small Monte Carlo cluster simulations colored by age for $log(age/yr)= 7.00, 7.50, 8.00, 8.50, 9.00$.  Here we show that young clusters of low mass can be found to reside in the region that the infinite mass limit predict for old stellar clusters. For a comparison with the observational data, the clusters from \citeauthor*{hunter2003} \citeyear{hunter2003} catalog are displayed as black dots. The clusters with photometric ages from \citeauthor*{santos} \citeyear{santos} (also presented in Table 1) are highlighted with pink squares. All the clusters presented in Table 2 are highlighted with yellow triangles. \normalsize}\label{fig:eight2}
\end{figure*}

Figures \ref{fig:one2} through \ref{fig:eight2} demonstrate the very difficult condition one faces in deriving unique ages of stellar clusters when mass is not given full consideration.  In creating the MASSCLEAN{\fontfamily{ptm}\selectfont \textit{colors}} database, we have created a distribution function of observed photometric properties with age and mass.  Our figures demonstrate just the UBV bands, however, our MASSCLEAN{\fontfamily{ptm}\selectfont \textit{colors}} database includes all photometric bands presently included in the Padova 2008 model release: UBVRIJHK.  While our figures seem to indicate there is little to match on, in fact, when simultaneously solving all three unique values of UBV ($U-B$, $B-V$ and $M_{V}$)
, there will be optimized, or most likely solutions when one searches in both age and mass grid space.  If additional photometric bands are available for a cluster, this can be further used to constrain the cluster properties.  The module which will provide such a search we call MASSCLEAN{\fontfamily{ptm}\selectfont \textit{age}}.

\section{Age Determination for Stellar Clusters -- MASSCLEAN{\fontfamily{ptm}\selectfont \textit{age}}}

The newest addition to the \texttt{MASSCLEAN} package (\citeauthor*{masscleanpaper} \citeyear{masscleanpaper}) is MASSCLEAN{\fontfamily{ptm}\selectfont \textit{age}}\footnote{As part of the \texttt{MASSCLEAN} package, MASSCLEAN{\fontfamily{ptm}\selectfont \textit{age}} is open source and freely available under GNU General Public License at: \url{http://www.physics.uc.edu/\textasciitilde popescu/massclean/}}, a program which uses the MASSCLEAN{\fontfamily{ptm}\selectfont \textit{colors}} database (\citeauthor*{paper2} \citeyear{paper2}) to determine the ages of stellar clusters.

Our program, MASSCLEAN{\fontfamily{ptm}\selectfont \textit{age}}, is based on a method analogous to $\chi^{2}$ minimization\footnote{$\chi^{2}$ minimization provides reliable results in the case of Gaussian distribution, so it did not apply to low and medium mass clusters.}. Based on the observational data, integrated magnitude and integrated colors, it searches for the minimum {\it hyper-radius} in the integrated magnitude, integrated colors, age, and mass hyperspace. This work specifically is based on a $M_{V}$, $(U-B)_{0}$, $(B-V)_{0}$, age, and mass hyperspace due to the availability of integrated $UBV$ colors and magnitudes from \citeauthor*{hunter2003} \citeyear{hunter2003}, but the code can provide results based on more colors. The most probable values for age and mass are reported, along with a list of other possible matches. The number of possible matches could be selected by the user or could be automatically computed based on the photometric errors. The results could be used to generate the confidence levels for age and mass (as presented in Tables 1 and 2 and Figure \ref{fig:nine2}), or to generate the age-mass probability distributions (as presented in Figure \ref{fig:ten2}).

By including additional bands (e.g. $(V-K)_{0}$) the accuracy in the age and mass determination would increase, and could minimize the degeneracies. In order to do it, a much larger version of the database is required, since the range of stochastic fluctuations is much larger in $(V-K)_{0}$ (\citeauthor*{paper2} \citeyear{paper2}).

\subsection{Age and Mass Determination for LMC Clusters}

\begin{figure}[htp]
\centering
\subfigure[]{\includegraphics[angle=270,width=8.25cm, bb=60 150 554 670]{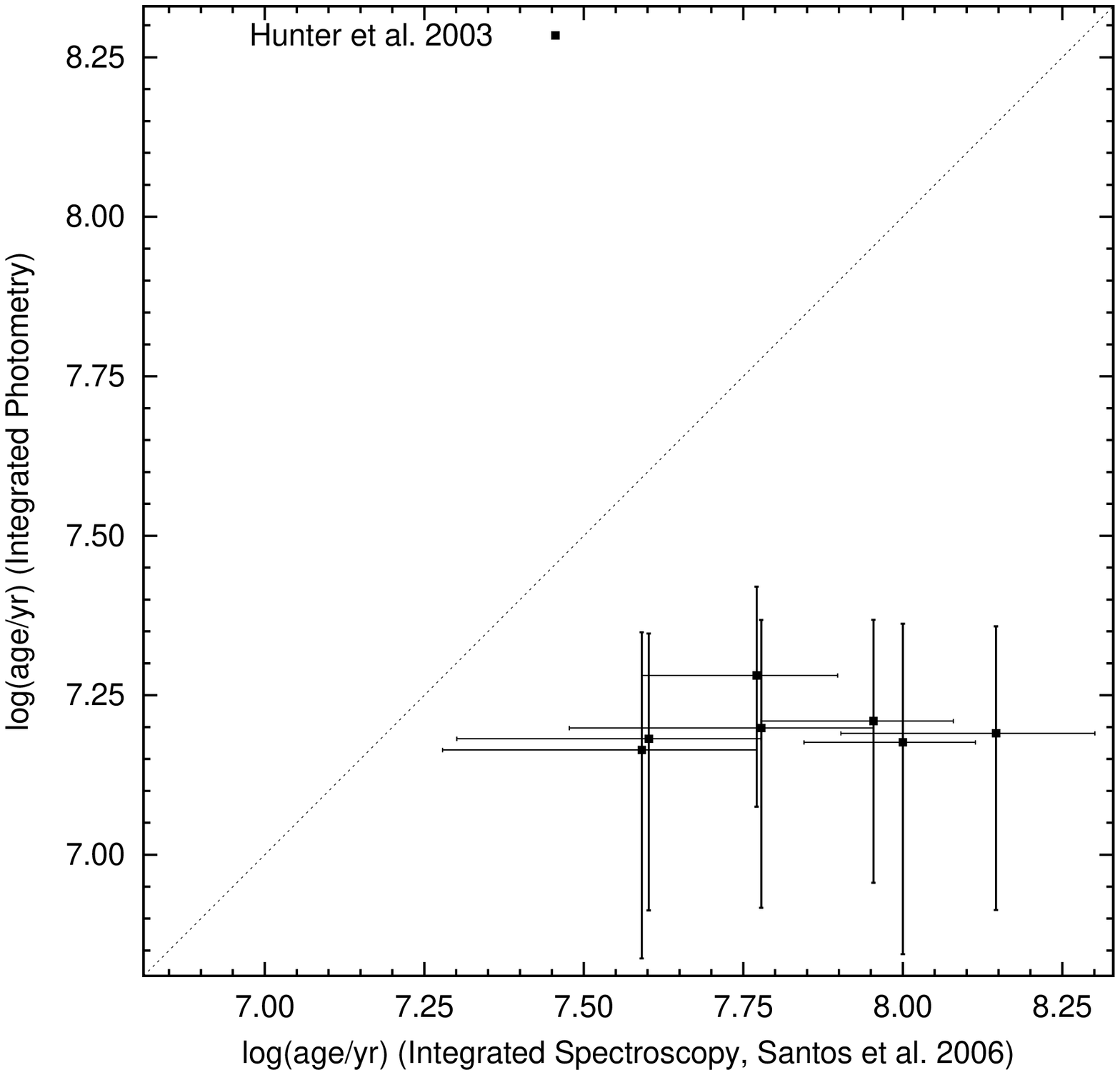}}
\subfigure[]{\includegraphics[angle=270,width=8.25cm, bb=60 150 554 670]{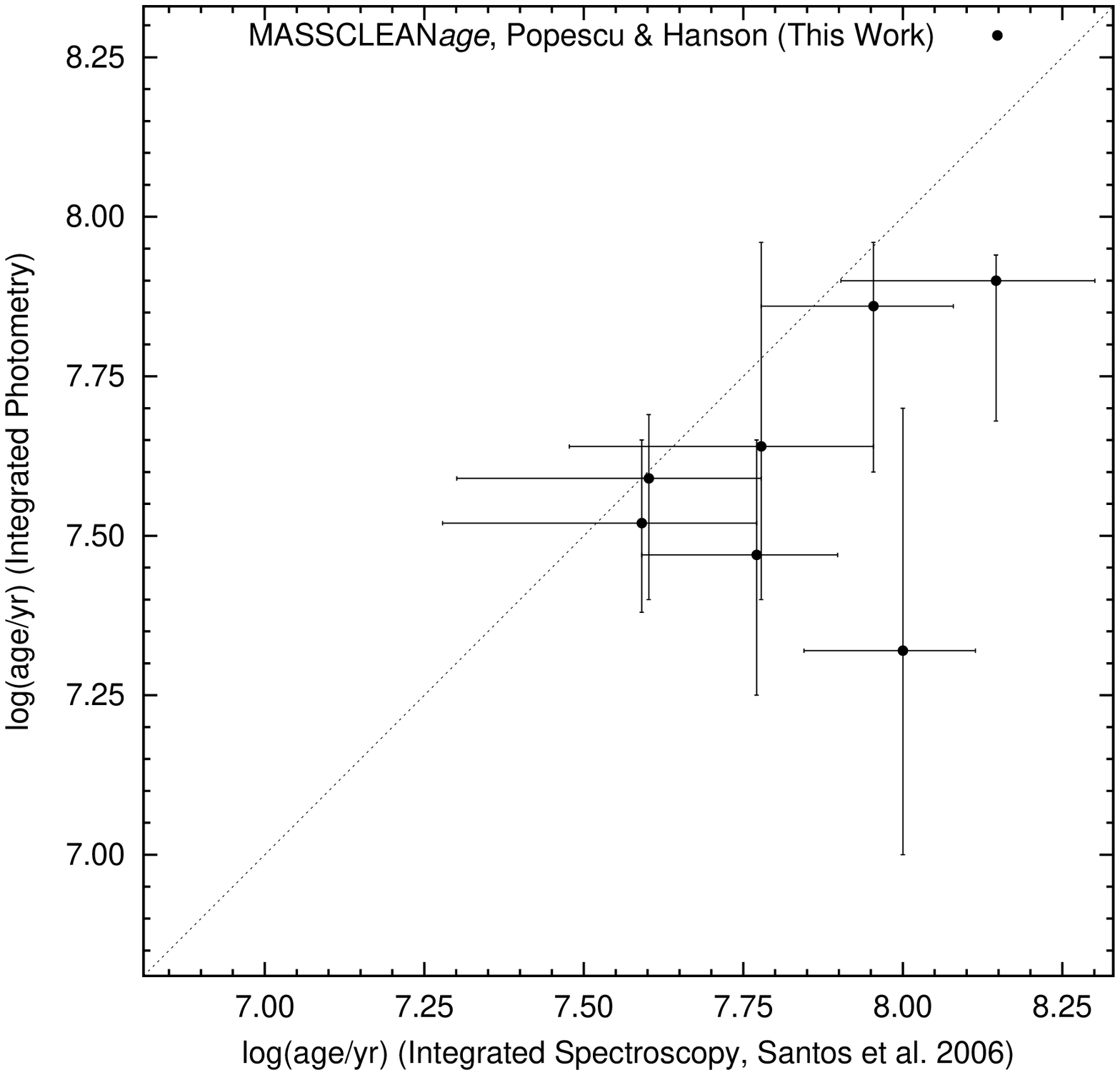}}
\caption{\small In (a) above, the $log(age/yr)$ as derived by Hunter et al.\ (2003) using UBV photometry alone is shown versus the $log(age/yr)$ as independently derived by \citeauthor*{santos} \citeyear{santos} based on integrated spectra of those same clusters.  In (b) below, we show the same set of stellar clusters with ages as derived by MASSCLEAN{\fontfamily{ptm}\selectfont \textit{age}} and using the same input UBV photometry as \citeauthor*{hunter2003} \citeyear{hunter2003} used. \normalsize}\label{fig:nine2}
\end{figure}

As a first demonstration of the MASSCLEAN{\fontfamily{ptm}\selectfont \textit{age}} code, we compare our values of ages for 7 clusters with both spectroscopic ages (\citeauthor*{santos} \citeyear{santos}) and photometric ages (\citeauthor*{hunter2003} \citeyear{hunter2003}). Our results are computed using the previously de-reddened values of $M_{V}$, $(U-B)_{0}$, and $(B-V)_{0}$ from the \citeauthor*{hunter2003} \citeyear{hunter2003} catalog. The data is presented in Table \ref{table:one}. The integrated photometry and age from \citeauthor*{hunter2003} \citeyear{hunter2003} is presented in the columns 2--5, and the spectroscopic ages from \citeauthor*{santos} \citeyear{santos} is presented in the column 6. The MASSCLEAN{\fontfamily{ptm}\selectfont \textit{age}} results are presented in the columns 7--9.  The same data is also presented in Figure \ref{fig:nine2} (a), (b). The integrated $(U-B)_{0}$ and $(B-V)_{0}$ colors are presented in Figure 8 as pink squares.

\begin{figure}[htp] 
\centering
\subfigure[]{\includegraphics[angle=270,width=8.0cm]{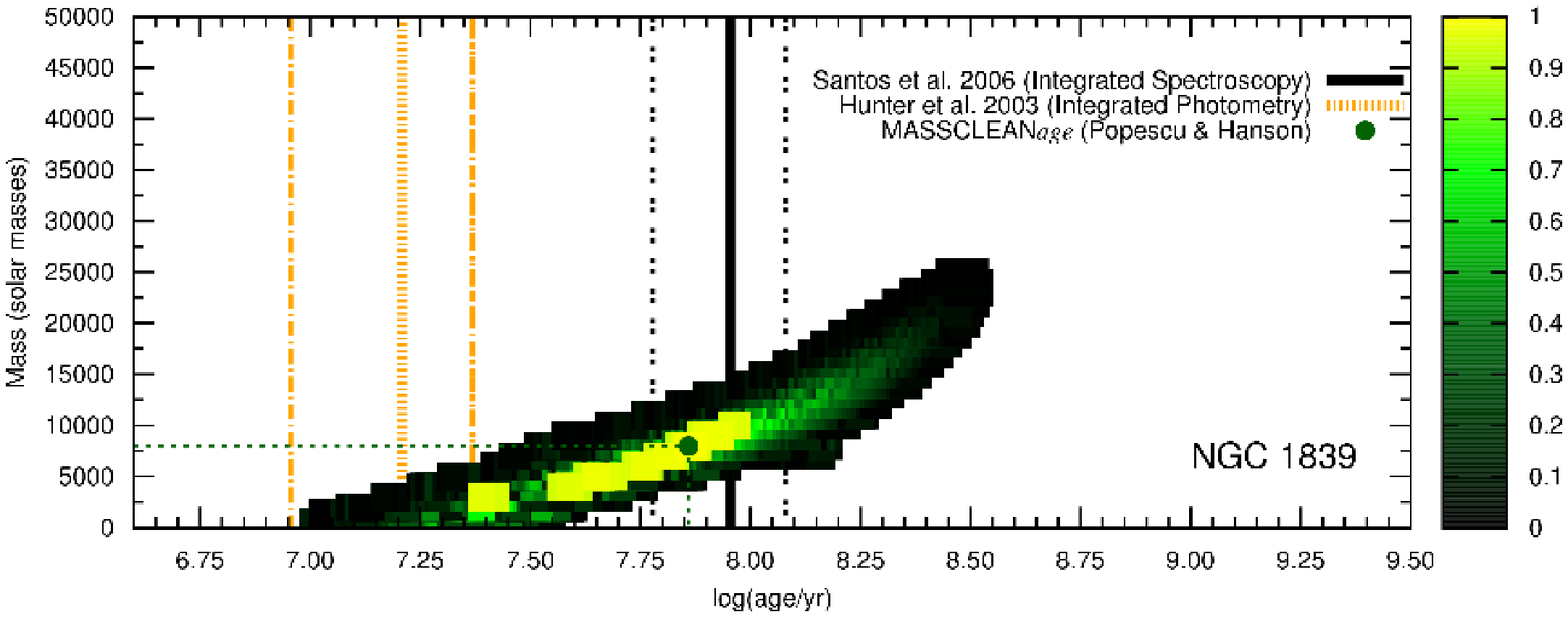}} 
\subfigure[]{\includegraphics[angle=270,width=8.0cm]{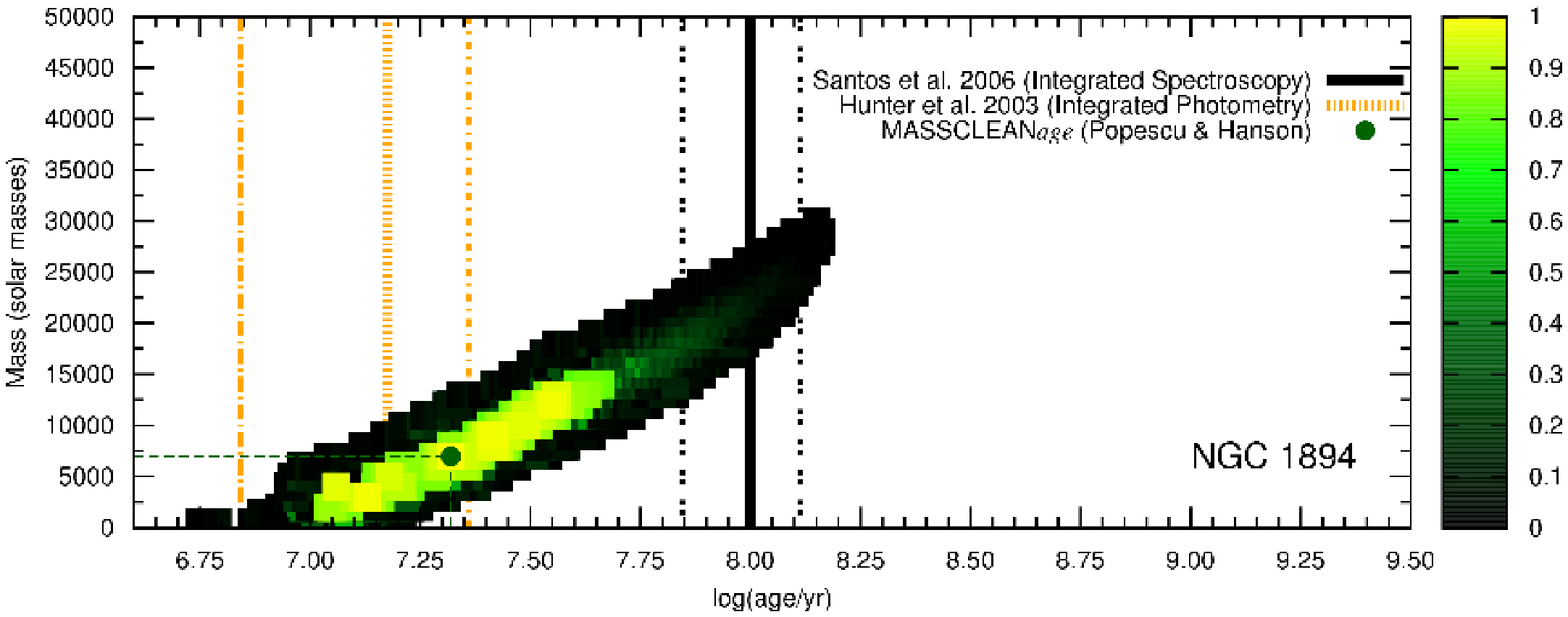}}
\subfigure[]{\includegraphics[angle=270,width=8.0cm]{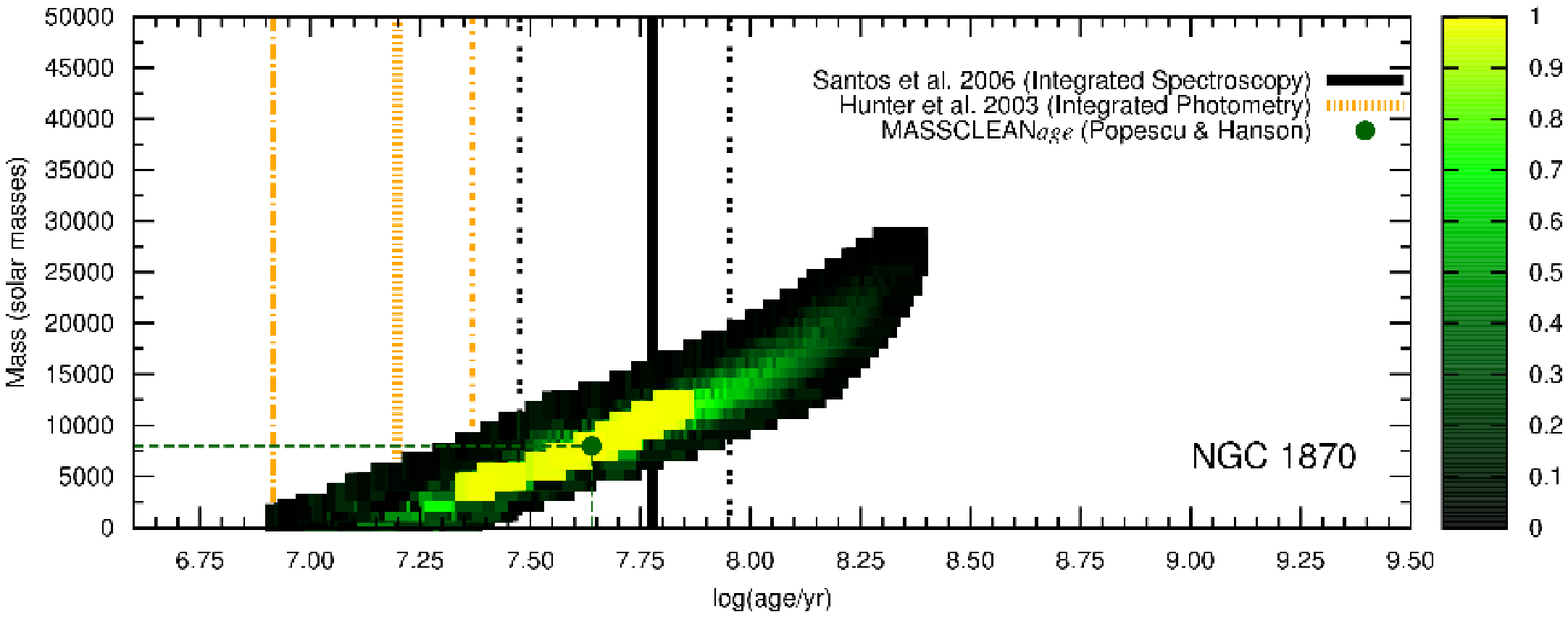}}
\caption{\small Output from MASSCLEAN{\fontfamily{ptm}\selectfont \textit{age}}: finding probabilistic regions in the mass - age plane, based on observed UBV magnitudes for these clusters.  These diagrams show the very strong correlation between cluster mass and age in integrated UBV colors.  Our best solution for NGC 1894 does not match the age given by \citeauthor*{santos} \citeyear{santos}.  \normalsize}\label{fig:ten2}
\end{figure}

The ages computed using MASSCLEAN{\fontfamily{ptm}\selectfont \textit{age}} are in good agreement with the spectroscopic ages from \citeauthor*{santos} \citeyear{santos}.
One object in Figure \ref{fig:nine2} still lies significantly away from the predicted age, as derived by \citeauthor*{santos} \citeyear{santos}, even using MASSCLEAN{\fontfamily{ptm}\selectfont \textit{age}}.  We investigated a few of the sources in this diagram by looking closely at the minimization solutions found within the MASSCLEAN{\fontfamily{ptm}\selectfont \textit{age}} hyperspace.  Absolutely every mass and age range available was specifically calculated against the observed UBV colors. This output was used to create the likelihood plots shown in Figure \ref{fig:ten2}.  These banana-shaped diagrams demonstrate the very strong degeneracy in UBV colors between age and mass.  
Here we see that for NGC 1894, which represents the most significantly desperate point found in Figure \ref{fig:nine2} (b), our best solution for the age does not match that derived by \citeauthor*{santos} \citeyear{santos} via integrated spectroscopy, but agrees with the age from \citeauthor*{hunter2003} \citeyear{hunter2003}. If the older \citeauthor*{santos} \citeyear{santos} age is correct, then this cluster is fairly massive ($\ge$ 15,000 $M_{\Sun}$ based on Figure \ref{fig:ten2}).

\begin{deluxetable}{lrrrrrrrrrr}
\tablecolumns{11}
\tablewidth{0pc}
\tablecaption{Results}
\tablehead{
\colhead{}    &  \multicolumn{4}{c}{{\footnotesize Integrated Photometry\tablenotemark{a}}} & \multicolumn{3}{c}{{\footnotesize Spectroscopy\tablenotemark{b} }} & \multicolumn{3}{c}{{\footnotesize MASSCLEAN}} \\
\cline{2-5} \cline{7-7} \cline{9-11}\\
\colhead{{\footnotesize Name}} & \colhead{{\footnotesize$M_{V}$}} & \colhead{{\footnotesize$(U-B)_{0}$}} & \colhead{{\footnotesize$(B-V)_{0}$}} & \colhead{{\footnotesize Age}}  & \colhead{{\tiny }} & \colhead{{\footnotesize Age}} & \colhead{{\tiny }} & \colhead{{\footnotesize Age}} & \colhead{{\footnotesize Age}} & \colhead{{\footnotesize Mass}} \\
\colhead{{\footnotesize }} & \colhead{{\footnotesize$(mag)$}} & \colhead{{\footnotesize$(mag)$}} & \colhead{{\footnotesize$(mag)$}} & \colhead{{\footnotesize $(Myr)$}}  & \colhead{{\tiny }} & \colhead{{\footnotesize $(Myr)$}} & \colhead{{\tiny }} & \colhead{{\footnotesize $(Myr)$}} & \colhead{{\footnotesize $(log)$}} & \colhead{{\footnotesize $(M_{\Sun}$)}} \\
\colhead{{\footnotesize$1$}} & \colhead{{\footnotesize$2$}} & \colhead{{\footnotesize$3$}} & \colhead{{\footnotesize$4$}} & \colhead{{\footnotesize$5$}} & \colhead{{\tiny }} & \colhead{{\footnotesize$6$}} & \colhead{{\tiny }} & \colhead{{\footnotesize$7$}} & \colhead{{\footnotesize$8$}} & \colhead{{\footnotesize$9$}}   }
\startdata
{\footnotesize NGC 1804} & {\footnotesize$-6.493$ \phn} & {\footnotesize$-0.523$ \phn} & {\footnotesize$-0.103$ \phn} & {\footnotesize$19.1\pm7.21$} & {\tiny } & {\footnotesize$60\pm20$} & {\tiny } & {\footnotesize$29.51^{+15.16}_{-11.73}$} & {\footnotesize$7.47^{+0.18}_{-0.22}$} & {\footnotesize$2,000^{+2,000}_{-1,300}$} \\

{\footnotesize NGC 1839} & {\footnotesize$-7.087$ \phn} & {\footnotesize$-0.347$ \phn} & {\footnotesize$0.002$ \phn} & {\footnotesize$16.2\pm7.16$} & {\tiny } & {\footnotesize$90\pm30$} & {\tiny } & {\footnotesize$72.44^{+18.76}_{-32.63}$} & {\footnotesize$7.86^{+0.10}_{-0.26}$} & {\footnotesize$8,000^{+2,000}_{-4,000}$} \\

{\footnotesize SL 237} & {\footnotesize$-6.951$ \phn} & {\footnotesize$-0.494$ \phn} & {\footnotesize$0.156$ \phn} & {\footnotesize$15.2\pm7.02$} & {\tiny } & {\footnotesize$40\pm20$} & {\tiny } & {\footnotesize$38.90^{+10.07}_{-13.78}$} & {\footnotesize$7.59^{+0.10}_{-0.19}$} & {\footnotesize$5,000^{+1,000}_{-3,000}$} \\

{\footnotesize NGC 1870} & {\footnotesize$-7.493$ \phn} & {\footnotesize$-0.423$ \phn} & {\footnotesize$0.082$ \phn} & {\footnotesize$15.8\pm7.54$} & {\tiny } & {\footnotesize$60\pm30$} & {\tiny } & {\footnotesize$43.65^{+23.96}_{-8.98}$} & {\footnotesize$7.64^{+0.19}_{-0.10}$} & {\footnotesize$8,000^{+4,000}_{-2,000}$} \\

{\footnotesize NGC 1894} & {\footnotesize$-7.956$ \phn} & {\footnotesize$-0.550$ \phn} & {\footnotesize$0.192$ \phn} & {\footnotesize$15.0\pm8.02$} & {\tiny } & {\footnotesize$100\pm30$} & {\tiny } & {\footnotesize$20.89^{+14.59}_{-1.40}$} & {\footnotesize$7.32^{+0.23}_{-0.03}$} & {\footnotesize$7,000^{+5,000}_{-1,000}$} \\

{\footnotesize NGC 1913} & {\footnotesize$-7.547$ \phn} & {\footnotesize$-0.493$ \phn} & {\footnotesize$0.257$ \phn} & {\footnotesize$14.6\pm7.72$} & {\tiny } & {\footnotesize$40\pm20$} & {\tiny } & {\footnotesize$33.11^{+11.55}_{-9.12}$} & {\footnotesize$7.52^{+0.13}_{-0.14}$} & {\footnotesize$7,000^{+2,000}_{-2,000}$} \\

{\footnotesize NGC 1943} & {\footnotesize$-7.255$ \phn} & {\footnotesize$-0.359$ \phn} & {\footnotesize$0.137$ \phn} & {\footnotesize$15.5\pm7.31$} & {\tiny } & {\footnotesize$140\pm60$} & {\tiny } & {\footnotesize$79.43^{+7.66}_{-31.57}$} & {\footnotesize$7.90^{+0.04}_{-0.22}$} & {\footnotesize$10,000^{+1,000}_{-4,000}$} \\

\enddata
\tablenotetext{a}{{\footnotesize Hunter et al. 2003}}
\tablenotetext{b}{{\footnotesize Santos et al. 2006}}
\label{table:one}
\end{deluxetable}

\subsection{Further Tests of the MASSCLEAN{\fontfamily{ptm}\selectfont \textit{age}} Program}

Regrettably, the \citeauthor*{santos} \citeyear{santos} study has only a small overlap with the \citeauthor*{hunter2003} \citeyear{hunter2003} catalog of clusters, a consistent, high quality UBV dataset with SSP derived ages.  But we wish to provide a further test of the MASSCLEAN{\fontfamily{ptm}\selectfont \textit{age}} package.  The \citeauthor*{hunter2003} \citeyear{hunter2003} study includes over 900 LMC clusters, but only a fraction of those clusters are good examples.  By this we mean, their location in the color-color diagram allows them to be more clearly aged using SSP models.

\begin{figure}[htp]
\centering
\includegraphics[angle=270,width=8.25cm, bb=60 150 554 670]{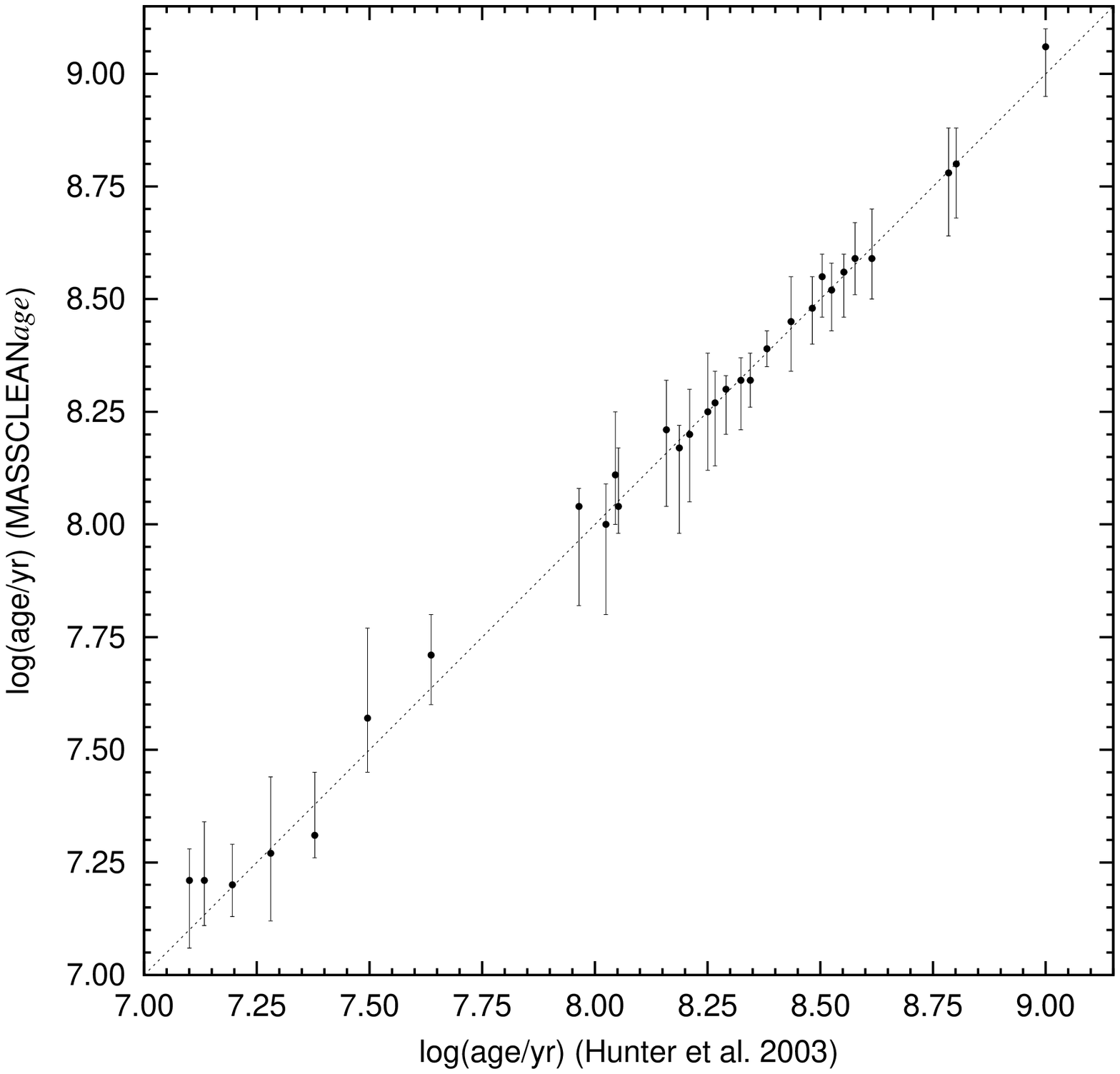}
\caption{\small We have further checked the consistency of the ages derived using MASSCLEAN{\it ages} by selecting a subset of \citeauthor*{hunter2003} \citeyear{hunter2003} clusters with previously well constrained ages.  These data are also shown in Table 2. \normalsize}\label{fig:eleven2}
\end{figure}

We selected a subset of 30 clusters, which cover a wide range of $(U-B)_{0}$ and $(B-V)_{0}$ colors, but are also located close to the infinite mass limit (see the yellow triangles in Figure \ref{fig:eight2}). The selected clusters are given in Table 2. Due to their locations in the color-color diagram, the age determination based on the classical $\chi^2$ minimization methods and infinite mass limit is expected to be relatively accurate (e.g.\ \citeauthor*{lancon2000} \citeyear{lancon2000}; \citeauthor*{paper2} \citeyear{paper2}). As shown in Figure \ref{fig:eleven2}, our MASSCLEAN{\fontfamily{ptm}\selectfont \textit{age}} results are in good agreement with the ages from \citeauthor*{hunter2003} \citeyear{hunter2003} based on SSP model fits. However, we can also derive the mass, and this is presented in column 8 of the Table 2. New values for age and mass for all the clusters with $UBV$ colors in the \citeauthor*{hunter2003} \citeyear{hunter2003} catalog and derived using MASSCLEAN{\fontfamily{ptm}\selectfont \textit{age}} will be presented in a subsequent paper (Popescu et al.\  2010, in preparation).

\begin{deluxetable}{lrrrrrrrr}
\tablecolumns{9}
\tablewidth{0pc}
\tablecaption{Results}
\tablehead{
\colhead{}    &  \multicolumn{4}{c}{{\footnotesize Integrated Photometry (Hunter et al. 2003)}} & \colhead{}  & \multicolumn{3}{c}{{\footnotesize MASSCLEAN}} \\
\cline{2-5} \cline{7-9}\\
\colhead{{\footnotesize Name}} & \colhead{{\footnotesize$M_{V}$}} & \colhead{{\footnotesize$(U-B)_{0}$}} & \colhead{{\footnotesize$(B-V)_{0}$}} & \colhead{{\footnotesize Age}}  & \colhead{{\tiny }}  & \colhead{{\footnotesize Age}} & \colhead{{\footnotesize Age}} & \colhead{{\footnotesize Mass}} \\
\colhead{{\footnotesize }} & \colhead{{\footnotesize$(mag)$}} & \colhead{{\footnotesize$(mag)$}} & \colhead{{\footnotesize$(mag)$}} & \colhead{{\footnotesize $(Myr)$}}  & \colhead{{\tiny }}  & \colhead{{\footnotesize $(Myr)$}} & \colhead{{\footnotesize $(log)$}} & \colhead{{\footnotesize $(M_{\Sun}$)}} \\
\colhead{{\footnotesize$1$}} & \colhead{{\footnotesize$2$}} & \colhead{{\footnotesize$3$}} & \colhead{{\footnotesize$4$}} & \colhead{{\footnotesize$5$}} & \colhead{{\tiny }} & \colhead{{\footnotesize$6$}} & \colhead{{\footnotesize$7$}} & \colhead{{\footnotesize$8$}}   }
\startdata
{\footnotesize KMK 88-88} & {\footnotesize$-7.848$ \phn} & {\footnotesize$-0.768$ \phn} & {\footnotesize$0.076$ \phn} & {\footnotesize$15.7$ \phn} & {\tiny }  & {\footnotesize$15.85^{+3.65 \phn}_{-2.36} \phn$} & {\footnotesize$7.20^{+0.09}_{-0.07}$} & {\footnotesize$7,000^{+3,000 \phn}_{-3,000}$} \\

{\footnotesize H 88-308} & {\footnotesize$-7.151$ \phn} & {\footnotesize$-0.716$ \phn} & {\footnotesize$0.265$ \phn} & {\footnotesize$13.6$ \phn} & {\tiny }  & {\footnotesize$16.22^{+5.66 \phn}_{-3.33} \phn$} & {\footnotesize$7.21^{+0.13}_{-0.10}$} & {\footnotesize$3,000^{+2,000 \phn}_{-500}$} \\

{\footnotesize NGC 2009} & {\footnotesize$-8.047$ \phn} & {\footnotesize$-0.735$ \phn} & {\footnotesize$0.204$ \phn} & {\footnotesize$12.6$ \phn} & {\tiny }  & {\footnotesize$16.22^{+2.84 \phn}_{-4.74} \phn$} & {\footnotesize$7.21^{+0.07}_{-0.15}$} & {\footnotesize$8,000^{+2,000 \phn}_{-5,000}$} \\

{\footnotesize SL 288} & {\footnotesize$-6.897$ \phn} & {\footnotesize$-0.562$ \phn} & {\footnotesize$-0.097$ \phn} & {\footnotesize$19.1$ \phn} & {\tiny }  & {\footnotesize$18.62^{+8.92 \phn}_{-5.44} \phn$} & {\footnotesize$7.27^{+0.17}_{-0.15}$} & {\footnotesize$1,500^{+2,500 \phn}_{-1,300}$} \\

{\footnotesize NGC 1922} & {\footnotesize$-7.742$ \phn} & {\footnotesize$-0.631$ \phn} & {\footnotesize$0.027$ \phn} & {\footnotesize$23.9$ \phn} & {\tiny }  & {\footnotesize$20.42^{+7.77 \phn}_{-2.22} \phn$} & {\footnotesize$7.31^{+0.14}_{-0.05}$} & {\footnotesize$6,000^{+5,000 \phn}_{-2,000}$} \\

{\footnotesize BSDL 1181} & {\footnotesize$-6.150$ \phn} & {\footnotesize$-0.203$ \phn} & {\footnotesize$0.050$ \phn} & {\footnotesize$31.3$ \phn} & {\tiny }  & {\footnotesize$37.15^{+21.73}_{-8.97} \phn$} & {\footnotesize$7.57^{+0.20}_{-0.12}$} & {\footnotesize$1,000^{+1,000 \phn}_{-300}$} \\

{\footnotesize SL 423} & {\footnotesize$-6.305$ \phn} & {\footnotesize$-0.442$ \phn} & {\footnotesize$0.071$ \phn} & {\footnotesize$43.3$ \phn} & {\tiny }  & {\footnotesize$51.29^{+11.81}_{-11.47} \phn$} & {\footnotesize$7.71^{+0.09}_{-0.11}$} & {\footnotesize$3,000^{+2,000 \phn}_{-1,000}$} \\

{\footnotesize HS 371} & {\footnotesize$-6.409$ \phn} & {\footnotesize$-0.292$ \phn} & {\footnotesize$0.166$ \phn} & {\footnotesize$105.8$ \phn} & {\tiny }  & {\footnotesize$100.00^{+23.03}_{-36.90} \phn$} & {\footnotesize$8.00^{+0.09}_{-0.20}$} & {\footnotesize$5,000^{+2,000 \phn}_{-2,000}$} \\

{\footnotesize KMK 88-45} & {\footnotesize$-5.116$ \phn} & {\footnotesize$-0.338$ \phn} & {\footnotesize$0.208$ \phn} & {\footnotesize$92.2$ \phn} & {\tiny }  & {\footnotesize$109.65^{+10.58}_{-43.58} \phn$} & {\footnotesize$8.04^{+0.04}_{-0.22}$} & {\footnotesize$2,000^{+500}_{-1,300 \phn}$} \\

{\footnotesize HS 318} & {\footnotesize$-4.462$ \phn} & {\footnotesize$-0.267$ \phn} & {\footnotesize$0.172$ \phn} & {\footnotesize$112.8$ \phn} & {\tiny }  & {\footnotesize$109.65^{+38.26}_{-14.15} \phn$} & {\footnotesize$8.04^{+0.13}_{-0.06}$} & {\footnotesize$700^{+800 \phn \phd \phn}_{-200}$} \\

{\footnotesize SL 408A} & {\footnotesize$-4.086$ \phn} & {\footnotesize$-0.277$ \phn} & {\footnotesize$0.208$ \phn} & {\footnotesize$111.0$ \phn} & {\tiny }  & {\footnotesize$128.82^{+49.00}_{-28.82} \phn$} & {\footnotesize$8.11^{+0.14}_{-0.11}$} & {\footnotesize$700^{+300 \phn \phd \phn}_{-400}$} \\

{\footnotesize HS 346} & {\footnotesize$-6.119$ \phn} & {\footnotesize$-0.228$ \phn} & {\footnotesize$0.211$ \phn} & {\footnotesize$153.9$ \phn} & {\tiny }  & {\footnotesize$147.91^{+18.05}_{-52.41} \phn$} & {\footnotesize$8.17^{+0.05}_{-0.19}$} & {\footnotesize$5,000^{+1,000 \phn}_{-2,000}$} \\

{\footnotesize KMK 88-76} & {\footnotesize$-4.673$ \phn} & {\footnotesize$-0.216$ \phn} & {\footnotesize$0.246$ \phn} & {\footnotesize$162.2$ \phn} & {\tiny }  & {\footnotesize$158.49^{+41.04}_{-46.29} \phn$} & {\footnotesize$8.20^{+0.10}_{-0.15}$} & {\footnotesize$1,500^{+500 \phn \phd \phn}_{-800}$} \\

{\footnotesize NGC 1695} & {\footnotesize$-6.845$ \phn} & {\footnotesize$-0.231$ \phn} & {\footnotesize$0.167$ \phn} & {\footnotesize$144.0$ \phn} & {\tiny }  & {\footnotesize$162.18^{+46.75}_{-52.53} \phn$} & {\footnotesize$8.21^{+0.11}_{-0.17}$} & {\footnotesize$11,000^{+4,000 \phn}_{-3,000}$} \\

{\footnotesize KMHK 494} & {\footnotesize$-4.378$ \phn} & {\footnotesize$-0.148$ \phn} & {\footnotesize$0.245$ \phn} & {\footnotesize$178.0$ \phn} & {\tiny }  & {\footnotesize$177.83^{+62.05}_{-46.00} \phn$} & {\footnotesize$8.25^{+0.13}_{-0.13}$} & {\footnotesize$1,000^{+500 \phn \phd \phn}_{-300}$} \\

{\footnotesize SL 110} & {\footnotesize$-5.334$ \phn} & {\footnotesize$-0.195$ \phn} & {\footnotesize$0.169$ \phn} & {\footnotesize$184.8$ \phn} & {\tiny }  & {\footnotesize$186.21^{+62.05}_{-46.00} \phn$} & {\footnotesize$8.27^{+0.07}_{-0.14}$} & {\footnotesize$3,000^{+1,000 \phn}_{-1,000}$} \\

{\footnotesize HS 218} & {\footnotesize$-5.607$ \phn} & {\footnotesize$-0.183$ \phn} & {\footnotesize$0.213$ \phn} & {\footnotesize$195.4$ \phn} & {\tiny }  & {\footnotesize$199.53^{+14.27}_{-41.04} \phn$} & {\footnotesize$8.30^{+0.03}_{-0.10}$} & {\footnotesize$4,000^{+1,000 \phn}_{-1,000}$} \\

{\footnotesize KMK 88-49} & {\footnotesize$-4.496$ \phn} & {\footnotesize$-0.160$ \phn} & {\footnotesize$0.198$ \phn} & {\footnotesize$210.8$ \phn} & {\tiny }  & {\footnotesize$208.93^{+14.27}_{-41.04} \phn$} & {\footnotesize$8.32^{+0.05}_{-0.11}$} & {\footnotesize$1,500^{+500 \phn \phd \phn}_{-500}$} \\

{\footnotesize KMK 88-35} & {\footnotesize$-4.488$ \phn} & {\footnotesize$-0.155$ \phn} & {\footnotesize$0.228$ \phn} & {\footnotesize$221.1$ \phn} & {\tiny }  & {\footnotesize$208.93^{+30.95}_{-26.96} \phn$} & {\footnotesize$8.32^{+0.06}_{-0.06}$} & {\footnotesize$1,500^{+500 \phn \phd \phn}_{-500}$} \\

{\footnotesize SL 160} & {\footnotesize$-5.077$ \phn} & {\footnotesize$-0.132$ \phn} & {\footnotesize$0.173$ \phn} & {\footnotesize$240.9$ \phn} & {\tiny }  & {\footnotesize$245.47^{+23.68}_{-21.60} \phn$} & {\footnotesize$8.39^{+0.04}_{-0.04}$} & {\footnotesize$3,000^{+500 \phn \phd \phn}_{-500}$} \\

{\footnotesize SL 580} & {\footnotesize$-4.957$ \phn} & {\footnotesize$-0.097$ \phn} & {\footnotesize$0.165$ \phn} & {\footnotesize$272.3$ \phn} & {\tiny }  & {\footnotesize$281.84^{+72.97}_{-63.06} \phn$} & {\footnotesize$8.45^{+0.10}_{-0.11}$} & {\footnotesize$3,000^{+1,000 \phn}_{-1,000}$} \\

{\footnotesize SL 224} & {\footnotesize$-5.489$ \phn} & {\footnotesize$-0.072$ \phn} & {\footnotesize$0.191$ \phn} & {\footnotesize$303.5$ \phn} & {\tiny }  & {\footnotesize$302.00^{+52.81}_{-50.80} \phn$} & {\footnotesize$8.48^{+0.07}_{-0.08}$} & {\footnotesize$5,000^{+1,000 \phn}_{-1,000}$} \\

{\footnotesize H 88-17} & {\footnotesize$-3.290$ \phn} & {\footnotesize$-0.051$ \phn} & {\footnotesize$0.209$ \phn} & {\footnotesize$335.1$ \phn} & {\tiny }  & {\footnotesize$331.13^{+49.06}_{-61.98} \phn$} & {\footnotesize$8.52^{+0.06}_{-0.09}$} & {\footnotesize$700^{+300 \phn \phd \phn}_{-200}$} \\

{\footnotesize HS 32} & {\footnotesize$-3.984$ \phn} & {\footnotesize$-0.044$ \phn} & {\footnotesize$0.172$ \phn} & {\footnotesize$319.3$ \phn} & {\tiny }  & {\footnotesize$354.81^{+43.29}_{-66.41} \phn$} & {\footnotesize$8.55^{+0.05}_{-0.09}$} & {\footnotesize$1,500^{+250 \phn \phd \phn}_{-500}$} \\

{\footnotesize H 88-235} & {\footnotesize$-4.046$ \phn} & {\footnotesize$-0.029$ \phn} & {\footnotesize$0.218$ \phn} & {\footnotesize$356.7$ \phn} & {\tiny }  & {\footnotesize$363.08^{+35.03}_{-74.67} \phn$} & {\footnotesize$8.56^{+0.04}_{-0.10}$} & {\footnotesize$1,500^{+500 \phn \phd \phn}_{-500}$} \\

{\footnotesize HS 76} & {\footnotesize$-4.331$ \phn} & {\footnotesize$0.010$ \phn} & {\footnotesize$0.272$ \phn} & {\footnotesize$411.6$ \phn} & {\tiny }  & {\footnotesize$389.05^{+112.14}_{-72.81}$} & {\footnotesize$8.59^{+0.11}_{-0.09}$} & {\footnotesize$2,000^{+500 \phn \phd \phn}_{-500}$} \\

{\footnotesize H 88-59} & {\footnotesize$-3.965$ \phn} & {\footnotesize$-0.008$ \phn} & {\footnotesize$0.258$ \phn} & {\footnotesize$377.6$ \phn} & {\tiny }  & {\footnotesize$389.05^{+78.69}_{-65.45} \phn$} & {\footnotesize$8.59^{+0.08}_{-0.08}$} & {\footnotesize$1,500^{+500 \phn \phd \phn}_{-500}$} \\

{\footnotesize SL 588} & {\footnotesize$-5.436$ \phn} & {\footnotesize$0.095$ \phn} & {\footnotesize$0.303$ \phn} & {\footnotesize$609.0$ \phn} & {\tiny }  & {\footnotesize$602.56^{+156.02}_{-166.04}$} & {\footnotesize$8.78^{+0.10}_{-0.14}$} & {\footnotesize$8,000^{+2,000 \phn}_{-3,000}$} \\

{\footnotesize HS 338} & {\footnotesize$-4.338$ \phn} & {\footnotesize$0.109$ \phn} & {\footnotesize$0.308$ \phn} & {\footnotesize$633.1$ \phn} & {\tiny }  & {\footnotesize$630.96^{+127.62}_{-152.33}$} & {\footnotesize$8.80^{+0.08}_{-0.12}$} & {\footnotesize$3,000^{+1,000 \phn}_{-1,000}$} \\

{\footnotesize SL 268} & {\footnotesize$-7.032$ \phn} & {\footnotesize$0.181$ \phn} & {\footnotesize$0.596$ \phn} & {\footnotesize$1,000.0$ \phn} & {\tiny }  & {\footnotesize$1,148.15^{+110.77}_{-256.90}$} & {\footnotesize$9.06^{+0.04}_{-0.11}$} & {\footnotesize$50,000^{+5,000}_{-10,000}$} \\

\enddata
\end{deluxetable}

\section{Discussion \& Conclusions}

That there would be degeneracies in the UBV color-color diagrams affecting the derivation of age as shown in Figures \ref{fig:seven2} and \ref{fig:eight2}, was previously recognized by many (e.g.\ \citeauthor*{bruzual2003} \citeyear{bruzual2003}; \citeauthor*{cervino2009} \citeyear{cervino2009}; \citeauthor*{buzzoni} \citeyear{buzzoni}; \citeauthor*{chiosi} \citeyear{chiosi}; \citeauthor*{bruzual2001} \citeyear{bruzual2001}; \citeauthor*{bruzual2010} \citeyear{bruzual2010}; \citeauthor*{cervino2004} \citeyear{cervino2004}; \citeauthor*{cervino2006} \citeyear{cervino2006}; \citeauthor*{fagiolini2007} \citeyear{fagiolini2007}; \citeauthor*{lancon2000} \citeyear{lancon2000}; \citeauthor*{lancon2009} \citeyear{lancon2009}; \citeauthor*{fouesneau} \citeyear{fouesneau}; \citeauthor*{fouesneau2} \citeyear{fouesneau2}; \citeauthor*{gonzalez} \citeyear{gonzalez}; \citeauthor*{gonzalez2005} \citeyear{gonzalez2005}; \citeauthor*{gonzalez2010} \citeyear{gonzalez2010}; \citeauthor*{jesus} \citeyear{jesus}; \citeauthor*{pandey} \citeyear{pandey}; \citeauthor*{paper2} \citeyear{paper2}).  
However, the scatter experienced in the integrated magnitude of a cluster 
also means using luminosity as a proxy for mass is highly dangerous for clusters of moderate and low mass.  This is clearly demonstrated in our simulations, particularly in Figure \ref{fig:four2}.  Present day SSP models are not designed to work with moderate and low mass clusters (e.g.\ \citeauthor*{bruzual2001} \citeyear{bruzual2001}; \citeauthor*{bruzual2010} \citeyear{bruzual2010}; \citeauthor*{lancon2000} \citeyear{lancon2000}; \citeauthor*{masscleanpaper} \citeyear{masscleanpaper}; \citeauthor*{paper2} \citeyear{paper2}). Yet, this is in fact the way in which mass functions for stellar clusters in other galaxies have been derived.  For some studies, it might be argued the masses are high enough that the errors are not extreme (\citeauthor*{larsen2000} \citeyear{larsen2000}), but this method is being extended to a lower mass regime such as in the LMC (e.g.\ \citeauthor*{billett} \citeyear{billett}; \citeauthor*{hunter2003} \citeyear{hunter2003}), simply because there was no other way to derive this critical information.

The MASSCLEAN{\fontfamily{ptm}\selectfont \textit{colors}} database is sufficiently populated now so it can be used to work back and derive mass and age for moderate and lower mass clusters in the LMC with a known (correctable) extinction.  To make this possible with just three input bands, the dataset 
uses the Padova 2008 models (\citeauthor*{padova2008} \citeyear{padova2008}) with a metallicity of $Z=0.008$ ([Fe/H]=-0.6).  While the simulations created for the database extend to ages up to $log(age/yr) = 9.5$, we have avoided applying the current inference package MASSCLEAN{\fontfamily{ptm}\selectfont \textit{age}} on clusters in the LMC for which previous indicators suggested ages greater than $log(age/yr) > 9.0$.   This is because of our concern in the evolutionary models at such age and the inevitable drop in metallicity for the very oldest LMC clusters.  It seems simplistic to assume the same metallicity can be applied to old clusters as to recently formed clusters, however, the metallicity of the LMC does appears to have been reasonably slow changing over a fairly long period of time, from relatively recent times until about $log(age/yr) = 9.5$ (\citeauthor*{bica} \citeyear{bica}; \citeauthor*{harris} \citeyear{harris}; \citeauthor*{piatti2009} \citeyear{piatti2009}).  These studies also show that for some young clusters, $log(age/yr) <  8$, the adopted metallicity for the current MASSCLEAN{\fontfamily{ptm}\selectfont \textit{colors}} database may be too low by a few dex. Whether the MASSCLEAN{\fontfamily{ptm}\selectfont \textit{colors}} database might be extended to include LMC clusters over a broader range of metallicities, from [Fe/H] = -1 to 0 and for $log(age/yr) > 9.5$, and if this might prove significant in deriving accurate age and mass estimates remains to be explored.   If such an additional parameter would be added to the database, more input photometry will be needed for the stellar cluster in question, to keep the search from becoming hopelessly degenerate over all the properties. 

\acknowledgements
We are grateful to suggestions made to an early draft of this work by Deidre Hunter and  Bruce Elmegreen.  Their ideas lead to significant improvements in the presentation. 
We thank the referee for useful comments and suggestions.  
This material is based upon work supported by the National Science Foundation under Grant No.\ 0607497 and more recently, Grant No.\ 1009550, to the University of Cincinnati. 




\begin{thebibliography}{}

\small \small
\bibitem[Bica et al. (1998)Bica et al.]{bica} Bica, E., Geisler, D., Dottori, H., Claria, J.H., Piatti, A.E. Santos, J.F. 1998, AJ, 116, 723
\bibitem[Billett et al. (2002)Billettt et al.]{billett} Billett, O. H., Hunter, D. A., Elmegreen, B. G. 2002, AJ, 123, 1454
\bibitem[Brocato et al. (1999)Brocato et al.]{brocato2000a} Brocato, E., Castellani, V., Raimondo, G., Romaniello, M. 1999, A\&AS, 136, 65
\bibitem[Brocato et al. (2000)Brocato et al.]{brocato2000b} Brocato, E., Castellani, V., Poli, F. M., Raimondo, G. 2000, A\&AS, 146, 91
\bibitem[Bruzual (2002)Bruzual]{bruzual2001} Bruzual, G. 2002, IAUS 207, 616
\bibitem[Bruzual (2010)Bruzual]{bruzual2010} Bruzual A., G. 2010, RSPTA, 368, 783
\bibitem[Bruzual \& Charlot (2003)Bruzual \& Charlot]{bruzual2003} Bruzual, G. \& Charlot, S. 2003, MNRAS, 344, 1000
\bibitem[Buzzoni (1989)Buzzoni]{buzzoni} Buzzoni, A. 1989, \apjs, 71, 817
\bibitem[Cantiello et al. (2003)Cantiello et al.]{cantiello} Cantiello, M., Raimondo, G., Brocato, E., Capaccioli, M. 2003, \aj, 125, 2783
\bibitem[Cervi\~no \& Luridiana (2004)Cervi\~no \& Luridiana]{cervino2004} Cervi\~no, M. \& Luridiana, V., 2004, A\&A, 413, 145
\bibitem[Cervi\~no \& Luridiana (2006)Cervi\~no \& Luridiana]{cervino2006} Cervi\~no, M. \& Luridiana, V., 2006, A\&A, 451, 475
\bibitem[Cervi\~no \& Valls-Gabaud (2003)Cervi\~no \& Valls-Gabaud]{cervino2003} Cervi\~no, M. \& Valls-Gabaud, D., 2003, MNRAS, 338, 481
\bibitem[Cervi\~no \& Valls-Gabaud (2009)Cervi\~no \& Valls-Gabaud]{cervino2009} Cervi\~no, M. \& Valls-Gabaud, D., 2009, Ap\&SS, 324, 91
\bibitem[Chandar et al. (2010a)Chandar et al.]{chandar2010a} Chandar, R., Fall, S.M, Whitmore, B.C. 2010a, ApJ, 711, 1263
\bibitem[Chandar et al. (2010b)Chandar et al.]{chandar2010b} Chandar, R., Whitmore, B.C., Fall, S.M. 2010b, ApJ, 713, 1343
\bibitem[Chiosi (1989)Chiosi]{chiosi} Chiosi, C. 1989, RMxAA, 18, 125
\bibitem[Fagiolini et al. (2007)Fagiolini et al.]{fagiolini2007} Fagiolini, M., Raimondo, G., Degl'Innocenti, S. 2007 A\&A, 462, 107
\bibitem[Fouesneau \& Lan{\c c}on (2009)Fouesneau \& Lan{\c c}on]{fouesneau} Fouesneau, M. \& Lan{\c c}on, A. 2009, preprint - astro-ph/0908.2742
\bibitem[Fouesneau \& Lan{\c c}on (2010)Fouesneau \& Lan{\c c}on]{fouesneau2} Fouesneau, M. \& Lan{\c c}on, A. 2010, preprint - astro-ph/1003.2334
\bibitem[Girardi et al. (1995)Girardi et al.]{girardi}Girardi, L., Bressan, A., Chiosi, C., Bertelli, G., Nasi, E. 1995 A\&A, 298, 87
\bibitem[Gonz\'alez et al. (2004)Gonz\'alez et al.]{gonzalez}Gonz\'alez, R.A., Liu, M.C., Bruzual A., G. 2004, \apj, 611, 270
\bibitem[Gonz\'alez-L\'opezlira et al. (2005)Gonz\'alez-L\'opezlira et al.]{gonzalez2005}Gonz\'alez-L\'opezlira, R.A., Albarr\'an, M.Y., Mouhcine, M., Liu, M.C., Bruzual-A., G., de Batz, B. 2005, MNRAS, 363, 1279
\bibitem[Gonz\'alez-L\'opezlira et al. (2010)Gonz\'alez-L\'opezlira et al.]{gonzalez2010}Gonz\'alez-L\'opezlira, R.A., Bruzual-A., G., Charlot, S., Ballesteros-Paredes, J., Loinard, L. 2010, MNRAS, preprint - astro-ph/0908.4133
\bibitem[Hancock et al. (2008)Hancock et al.]{hancock} Hancock, M., Smith, B.J., Giroux, M.L., Struck, C. 2008, MNRAS 389, 1470
\bibitem[Harris \& Zaritsky (2009)Harris \& Zaritsky]{harris} Harris, J. \& Zaritsky, D. 2009, AJ, 138, 1243
\bibitem[Hunter et al. (2003)Hunter et al.]{hunter2003} Hunter, D.A., Elmegreen, B.G., Dupuy, T.J., Mortonson, M. 2003, \aj, 126, 1836
\bibitem[Kroupa (2002)Kroupa]{Kroupa2002} Kroupa, P. 2002, {\it Sci}, 295, 82
\bibitem[Lamers et al.(2005)Lamers et al.]{lamers2005}Lamers, H. J. G. L. M., Gieles, M., Bastian, N., Baumgardt, H., Kharchenko, N. V., Portegies Zwart, S. 2005, A\&A, 441, 117
\bibitem[Lan{\c c}on \& Mouhcine (2000)Lan{\c c}on \& Mouhcine]{lancon2000} Lan{\c c}on, A. \& Mouhcine, M. 2000, ASPC, 211, 34
\bibitem[Lan{\c c}on \& Mouhcine (2002)Lan{\c c}on \& Mouhcine]{lancon2002} Lan{\c c}on, A. \& Mouhcine, M. 2002, A\&A, 393, 167
\bibitem[Lan{\c c}on \& Fouesneau (2009)Lan{\c c}on \& Fouesneau]{lancon2009} Lan{\c c}on, A. \& Fouesneau, M. 2009, ASPC, preprint - astro-ph/0903.4557
\bibitem[Larsen \& Richtler (2000)Larsen \& Richtler]{larsen2000} Larsen, S. S. \& Richtler, T. 2000, A\&A, 354, 836
\bibitem[Larsen (2009)Larsen]{larsen2009}Larsen, S.S. 2009, A\&A, 494, 539
\bibitem[Lejeune \& Schaerer(2001)Lejeune \& Schaerer]{geneva1} Lejeune, T. \& Schaerer, D. 2001, \aap, 366, 538
\bibitem[Ma\'iz Apell\'aniz (2009)Ma\'iz Apell\'aniz]{jesus} Ma\'iz Apell\'aniz, J. 2009, \apj, 699, 1938
\bibitem[Marigo et al.(2008)Marigo et al.]{padova2008} Marigo, P., Girardi, L., Bressan, A., Groenewegen, M. A. T., Silva, L., Granato, G. L. 2008, A\&A, 482, 833
\bibitem[Mora et al.(2009)Mora et al.]{mora2009} Mora, M.D. Larsen, S.S., Kissler-Patig, M., Brodie, J.O., Richtler, T. 2009 A\&A, 501, 949
\bibitem[Pandey et al.(2010)Pandey et al.]{pandey} Pandey, A.K., Sandhu, T.S., Sagar, R., \& Battinelli, P. 2010, MNRAS, 403, 1491
\bibitem[Pessev et al. (2009)Pessev et al.]{pessev} Pessev, P. M., Goudfrooij, P., Puzia, T. H., Chandar, R. 2008, MNRAS, 385, 1535
\bibitem[Piatti et al. (2009)Piatti et al.]{piatti2009}Piatti, A.E., Geisler, D., Sarajedini, A., Gallart, C. 2009, A\&A, 501, 585
\bibitem[Popescu \& Hanson (2009)Popescu \& Hanson]{masscleanpaper} Popescu, B. \& Hanson, M.M. 2009, \aj, 138, 1724
\bibitem[Popescu \& Hanson (2010a)Popescu \& Hanson]{paper2} Popescu, B. \& Hanson, M.M. 2010a,\apj, 713, L21 
\bibitem[Popescu \& Hanson (2010b)Popescu \& Hanson]{iaus266} Popescu, B. \& Hanson, M.M. 2010b, Proceedings International Astronomical Union Symposium No. 266, p. 511, {\it Star Clusters - Basic Galactic Building Blocks throughout Time and Space}, Editors: Richard de Grijs \& Jacques R. D. Lepine 
\bibitem[Raimondo et al. (2005)Raimondo et al.]{raimondo2005} Raimondo, G., Brocato, E., Cantiello, M., Capaccioli, M. 2005, \aj, 130, 2625
\bibitem[Raimondo (2009)Raimondo]{raimondo2009} Raimondo, G. 2009, \apj, 700, 1247
\bibitem[Santos \& Frogel (1997)Santos \& Frogel]{santos} Santos, J.F.C. \& Frogel, J.A. 1997, \apj, 479, 764 
\bibitem[Santos et al.(2006)Santos et al.]{santos} Santos, J.F.C, Clari\'a, J.J., Ahumada, A.V., Bica, E., Piatti, A.E., Parisi, M.C. 2006, A\&A, 448, 1023
\bibitem[Searle et al. (1980)Searle et al.]{searle} Searle, L., Wilkinson, A., Bagnuolo, W.G. 1980, \apj, 239, 803
\bibitem[van den Bergh(1981)van den Bergh]{vdb}van den Bergh, S. 1981, A\&AS, 46, 79
\normalsize
\end{thebibliography}
\end{document}